\documentclass[11pt,a4paper]{article}
\pdfoutput=1

\usepackage{jcappub}
\usepackage{graphicx}
\usepackage{amssymb}
\usepackage{epstopdf}
\usepackage{mathtools}
\usepackage{color}
\usepackage[section]{placeins}
\usepackage[usenames,dvipsnames]{xcolor}
\usepackage{epsfig,amssymb,bm, graphicx, amsmath}
\usepackage[normalem]{ulem}
\usepackage{enumerate,enumitem}

\title{\center{\huge{Rolling in the Modulated Reheating Scenario}}}
\author[a,b]{Naoya Kobayashi,}
\author[b,c]{Takeshi Kobayashi,}
\author[b,c,d]{Adrienne L. Erickcek}
\affiliation[a]{Department of Astronomy and Astrophysics, University of Toronto,\\
50 St. George Street, Toronto, Ont. M5S 3H4, Canada}
\affiliation[b]{Canadian Institute for Theoretical Astrophysics, University of Toronto,\\
60 St. George Street, Toronto, Ont. M5S 3H8, Canada}
\affiliation[c]{Perimeter Institute for Theoretical Physics,\\
31 Caroline Street North, Waterloo, Ont. N2L 2Y5, Canada}
\affiliation[d]{
Department of Physics and Astronomy, University of North Carolina, Chapel Hill, Philips Hall CB 3255, Chapel Hill, NC 27599 USA}
\emailAdd{nkobayas@cita.utoronto.ca}
\emailAdd{takeshi@cita.utoronto.ca}
\emailAdd{erickcek@physics.unc.edu}
\abstract{In the modulated reheating scenario, the field that drives inflation has a spatially varying decay rate, and the resulting inhomogeneous reheating process generates adiabatic perturbations. We examine the statistical properties of the density perturbations generated in this scenario. Unlike earlier analyses, we include the dynamics of the field that determines the inflaton decay rate. We show that the dynamics of this modulus field can significantly alter the amplitude of the power spectrum and the bispectrum, even if the modulus field has a simple potential and its effective mass is smaller than the Hubble rate.  In some cases, the evolution of the modulus amplifies the non-Gaussianity of the perturbations to levels that are excluded by recent observations of the cosmic microwave background. Therefore, a proper treatment of the modulus dynamics is required to accurately calculate the statistical properties of the perturbations generated by modulated reheating. }

\begin{document}
\maketitle
\flushbottom

\section{Introduction}
Inflation provides a mechanism to generate the small density perturbations that seed the growth of all cosmic structures\cite{guth81, mukhanov81, sato80, guth82, hawking82, linde82, starobinsky82, bardeen83}. In the simplest single-field inflation models, these perturbations originate from quantum fluctuations in the field that dominates the energy density of the early Universe and drives inflation.  However, there may be multiple degrees of freedom in the early Universe in addition to the dominant inflaton field, and these spectator fields may also contribute to the production of primordial density perturbations. 

The modulated reheating scenario\cite{dvali04, kofman03} provides an alternative perturbation-generating mechanism by considering such additional degrees of freedom. Inflation ends when the inflaton starts oscillating about the minimum of its potential. Given that the oscillations are harmonic, the inflaton's energy density behaves like non-relativistic matter. The inflaton then decays during a period called reheating, during which the Universe transitions to a radiation-dominated state. In the modulated reheating scenario, the decay rate of the inflaton is a function of one or more modulus fields that have energy densities that are much smaller than the energy density of the inflaton field. The modulus fields are assumed to have effective masses that are small compared to the inflationary Hubble rate, and thus they acquire nearly scale-invariant fluctuations during inflation. Inhomogeneities in the modulus fields then modulate the inflaton decay rate, introducing spatial variations in the time of reheating. Since matter and radiation densities redshift at different rates, the inhomogeneous transition from an oscillating-inflaton-dominated Universe to a radiation-dominated Universe leads to the generation of adiabatic density fluctuations.

We can discriminate between the modulated reheating scenario and single-field inflation by studying the power spectrum and higher-order statistics of the density perturbations. It has been shown previously that the non-Gaussianity of the density perturbations generated by modulated reheating can be much larger than the non-Gaussianity generated in single-field inflation\cite{zaldarriaga04, bartolo04, vernizzi04, suyama08, ichikawa08, elliston13}. Since the modulus field must be light during inflation in order to acquire fluctuations, it was thought that Hubble friction would prevent the modulus from significantly rolling prior to reheating.  Consequently, these analyses neglected the dynamics of the modulus field when computing the statistical properties of the perturbations generated by modulated reheating.\footnotemark\footnotetext{Refs.\cite{zaldarriaga04, suyama08, elliston13} include modulus dynamics during inflation, but not after inflation ends.  We will show that the evolution of the modulus is more important after inflation because it leads to a time-varying decay rate and also because Hubble friction becomes weaker while the inflaton field oscillates.} 
Our goal is to improve the calculations of the amplitude of the power spectrum $\mathcal{P}_{\zeta}$ and the non-linearity parameter $f_{\rm{NL}}$ by fully accounting for the evolution of the modulus field.  We will show that modulus evolution can significantly alter the statistical properties of the resulting curvature perturbation, even if the modulus mass is small compared to the Hubble rate.

Unlike single-field models, for which the local-type $|f_{\rm{NL}}|$ parameter cannot be greater than $\sim1$ \cite{allen87, falk92, gangui94, maldacena03, acquaviva03, seery05}, some modulated reheating models produce perturbations with significant deviations from Gaussianity in the local limit\cite{zaldarriaga04, bartolo04, vernizzi04, suyama08, ichikawa08, elliston13}.  To demonstrate that the evolution of the modulus can affect the level of non-Gaussianity, we will present models of modulated reheating that have $|f_{\rm{NL}}|\lesssim5$ when the modulus evolution is neglected, while the modulus evolution generates $|f_{\rm{NL}}| \gtrsim \mathcal{O}(10)$. This modification is especially significant given the tight constraints on non-Gaussianity recently provided by the Planck satellite\cite{ade13no2}:
\begin{equation} \label{2.7}
f_{\rm{NL}}=2.7\pm5.8\phantom{a}(68\%\text{CL}).
\end{equation}
The amplitude of the power spectrum $\mathcal{P}_{\zeta}$ is also greatly affected by the modulus dynamics. Neglecting the modulus's evolution can lead to an overestimation of $\mathcal{P}_{\zeta}$ by orders of magnitude. We will show that there are two general cases in which the modulus-rolling corrections to $\mathcal{P}_{\zeta}$ and $f_{\rm{NL}}$ can be significant: when the inflaton decay rate rapidly changes during reheating, and when the effective mass of the modulus changes between inflation and reheating. We will provide specific case studies that exemplify these two scenarios and demonstrate how neglecting the evolution of the modulus significantly changes the perturbations generated in these models. Unfortunately, it is not possible to evaluate how these corrections affect previous analyses of the modulated reheating scenario\cite{zaldarriaga04, bartolo04, vernizzi04, suyama08, ichikawa08, elliston13}; since these analyses assumed that the modulus's evolution is unimportant, they did not specify the modulus potential. Depending on the choice of the modulus potential, the results derived in these previous works could be significantly altered when the evolution of the modulus is included. In order to utilize statistics of the curvature perturbations to probe the modulated reheating mechanism, it is crucial that one is aware of the size of the corrections that result from modulus dynamics.  To that end, we will establish tests that determine the importance of the modulus's evolution given functional forms for the modulus potential and inflaton decay rate.

In section 2, we derive analytic expressions for the power spectrum and the non-linearity parameter while fully taking the modulus's evolution into account. In section 3, we examine models in which the effective mass of the modulus is constant, but the inflaton decay rate rapidly changes during reheating.   We consider the opposite scenario in section 4 by analyzing a model with a nearly constant decay rate and a varying effective modulus mass.
In both cases, we will show that the modulus's evolution can have a significant impact on the power spectrum and the non-linearity parameter. The analytic formulas derived in section 2 are also compared to numerical solutions in order to test the validity of the simplifying assumptions we used in their derivations. We discuss our findings and conclude in section 5.

\section{Analytic Treatment of the Curvature Perturbations}
\subsection{A General Expression for $\mathcal{P}_{\zeta}$ and $f_{\rm{NL}}$}
The purpose of this section is to obtain analytical expressions for the power spectrum and the bispectrum of the curvature fluctuations from modulated reheating without disregarding the rolling of the modulus. For simplicity we take the inflaton decay rate $\Gamma=\Gamma(\sigma)$ to be controlled by only one modulus field ($\sigma$) that has a generic potential $V(\sigma)$.\footnotemark\footnotetext{Our formalism can be easily generalized to accommodate inflaton decay rates that depend on multiple modulus fields; see Ref.\cite{suyama08} for an example of this extension.}   We define the power spectrum $\mathcal{P}_{\zeta}=(k^3/2\pi^2)P_{\zeta}$ and the non-linearity parameter $f_{\rm{NL}}$ for the gauge-invariant curvature perturbation $\zeta$ by 
\begin{equation} \label{2pt}
\langle\zeta(\vec{k}_1)\zeta(\vec{k}_2)\rangle = (2\pi)^3P_{\zeta}(k_1)\delta^{(3)}(\vec{k}_1+\vec{k}_2),
\end{equation}
\begin{equation} \label{3pt}
\langle\zeta(\vec{k}_1)\zeta(\vec{k}_2)\zeta(\vec{k}_3)\rangle = (2\pi)^3\frac{6}{5}f_{\rm{NL}}\left[P_{\zeta}(k_1)P_{\zeta}(k_2)+P_{\zeta}(k_2)P_{\zeta}(k_3)+P_{\zeta}(k_3)P_{\zeta}(k_1)\right]\delta^{(3)}(\vec{k}_1+\vec{k}_2+\vec{k}_3).
\end{equation}
In Eq.(\ref{3pt}), we have implicitly assumed that the bispectrum has a
local shape, which is the type of non-Gaussianity mainly produced by modulated reheating.

In the modulated reheating scenario, density perturbations are continuously generated until the transition from an oscillatory-inflaton-dominated universe to a radiation-dominated universe is complete. To obtain the final value of $\zeta$, we therefore need to evolve these perturbations to some time after reheating is completed. We use the $\delta\mathcal{N}$-formalism\cite{starobinsky85, sasaki96, wands00, lyth05} in which the curvature perturbation on a uniform-energy-density hypersurface at $t_f$ is given by
\begin{equation} \label{zeta}
\zeta(t_f,\vec{x})=\mathcal{N}(t_f,t_i,\vec{x}) - \langle \mathcal{N}(t_f,t_i,\vec{x}) \rangle,
\end{equation}
where the slicing at the initial time~$t_i$ is a flat slicing,
$\mathcal{N}(t_f,t_i,\vec{x}) \equiv \int_{t_i}^{t_f}H(t,\vec{x})dt$ is the e-folding number, and $\langle\phantom{a}\rangle$ indicates spatial average. 

The curvature perturbation that originates from fluctuations in some field $\sigma$ can be written as 
\begin{equation}
\zeta(t_f,\vec{x}) = \frac{\partial\mathcal{N}}{\partial\sigma}\delta\sigma+\frac{1}{2}\frac{\partial^2\mathcal{N}}{\partial\sigma^2}(\delta\sigma^2-\langle\delta\sigma^2\rangle)+\mathcal{O}(\delta\sigma^3).
\end{equation}
By Fourier transforming this equation, one can obtain the power spectrum $\mathcal{P}_{\zeta}$ and the non-linearity parameter $f_{\rm{NL}}$ from Eqs.(\ref{2pt}, \ref{3pt}):
\begin{equation} \label{P1} 
\mathcal{P}_{\zeta}=\left(\frac{\partial\mathcal{N}\hfill}{\partial\sigma}\right)^2\mathcal{P}_{\delta\sigma}, 
\end{equation}
\begin{equation} \label{fNL1}
f_{\rm{NL}}=\frac{5}{6}\left(\frac{\partial\mathcal{N}}{\partial\sigma}\right)^{-2}\frac{\partial^2\mathcal{N}}{\partial\sigma^2}.
\end{equation}
The total curvature perturbation in the modulated reheating scenario
comes from both the process of inhomogeneous reheating and the quantum
fluctuations of the inflaton field. 
However, the purpose of this paper is to study how the statistical
properties of $\zeta$ are affected by the dynamics of the modulus field
$\sigma$.  
Assuming that the energy density of the modulus field stays subdominant to the total
energy density before reheating, the curvature perturbation generated by the inflaton field is independent of the modulus dynamics.\footnotemark\footnotetext{To be precise, the modulus does not affect the inflaton-induced perturbations produced during inflation. However, in the presence of the modulus, the inflaton field fluctuations can further generate curvature perturbations in the post-inflationary era by affecting the modulus's evolution \cite{Kobayashi:2013awa}.}
We therefore drop the contribution to the total $\zeta$ from the inflaton\footnotemark\footnotetext{See Ref.\cite{ichikawa08} for a study including contributions from both the inflaton and modulated reheating with a static modulus.} for simplicity, and we proceed with the power spectrum and the non-linearity parameter as given in Eqs.(\ref{P1}, \ref{fNL1}).

To obtain $\mathcal{N}=\mathcal{N}(t_f,t_*,\vec{x})$, we take $t_f$ to be any time after the inflaton energy density is converted into radiation. Assuming that the modulus energy density stays subdominant, the curvature perturbation is conserved on superhorizon scales after the Universe becomes radiation-dominated, because the perturbations generated during inhomogeneous reheating are adiabatic. Next, we take $t_i=t_*$ to be the time when a scale of interest crosses the horizon, given by the condition $k=aH$ where $k$ is the comoving wave number. The importance of this choice of $t_i$ is two-fold. First, all light fields acquire fluctuations with $\mathcal{P}_{\delta\sigma_*}\simeq(H_*/2\pi)^2$ at horizon crossing. Secondly, assuming that $\sigma$ has weak self-interactions, we can take $\delta\sigma_*$ to be Gaussian: $\langle\delta\sigma_*(\vec{k}_1)\delta\sigma_*(\vec{k}_2)\delta\sigma_*(\vec{k}_3)\rangle=0$. We cannot neglect the three-point function of $\delta\sigma$ at later times (e.g. at reheating) because $\delta\sigma$ will have a non-Gaussian component due to the non-linear evolution of $\sigma$. 

To proceed further, we assume that the Universe transitions from the inflaton-oscillation phase to the radiation-dominated phase instantaneously. We define the time of sudden inflaton decay $t_{\rm{reh}}$ by $\Gamma(\sigma(t_{\rm{reh}}))/H(t_{\rm{reh}})=\beta$, where we take $\beta$ to be a constant of order unity. Then, using the fact that the Universe is effectively dominated by matter from the end of inflation to reheating and dominated by radiation after reheating, we obtain
\begin{align} \label{efold2}
\mathcal{N}(t_f,t_*,\vec{x})&= \int_{t_*}^{t_{\rm{end}}}Hdt+\int_{\rho_{\rm{end}}}^{\rho_{f}}\frac{H}{\dot{\rho}}d\rho \nonumber \\ &=\mathcal{N}_*+\int_{\rho_{\rm{end}}}^{\rho_{\rm{reh}}}\frac{d\rho_{\phi}}{-3\rho_{\phi}} +\int_{\rho_{\rm{reh}}}^{\rho_{f}}\frac{d\rho_\gamma}{-4\rho_\gamma} \nonumber \\ 
&= \mathcal{N}_*-\frac{1}{2}\ln\left[\frac{H(t_f)}{H(t_{\rm{end}})}\right] - \frac{1}{6}\ln\left[\frac{\Gamma(\sigma_{\rm{reh}})}{\beta H(t_{\rm{end}})}\right]
\end{align}
where $\phi$ is the inflaton field (which behaves like matter after inflation), $\gamma$ stands for radiation, and subscripts $\{*, \rm{end},\rm{reh}\}$ indicate evaluation at horizon crossing, end of inflation, and time of instant inflaton decay (or equivalently, reheating), respectively. We make one exception and define the e-folding number between horizon crossing and the end of inflation by $\mathcal{N}_*\equiv\int_{t_*}^{t_{\rm{end}}}Hdt$. We will follow these conventions for the rest of this paper. 

From Eq.(\ref{efold2}) we see that the e-folding number depends on the modulus only through $\Gamma(\sigma_{\rm{reh}})$, so we obtain
\begin{equation} \label{n'n''}
\frac{\partial\mathcal{N}\hfill}{\partial\sigma_{\rm{reh}}}=\frac{\partial\mathcal{N}\hfill}{\partial\Gamma_{\rm{reh}}}\Gamma'(\sigma_{\rm{reh}})= -\frac{1}{6}\frac{\Gamma'(\sigma_{\rm{reh}})}{\Gamma(\sigma_{\rm{reh}})}, \phantom{hihi}\frac{\partial^2\mathcal{N}\hfill}{\partial\sigma_{\rm{reh}}^2}= -\frac{1}{6}\frac{\Gamma''(\sigma_{\rm{reh}})}{\Gamma(\sigma_{\rm{reh}})} + \frac{1}{6}\left(\frac{\Gamma'(\sigma_{\rm{reh}})}{\Gamma(\sigma_{\rm{reh}})}\right)^2,
\end{equation}
where a prime indicates differentiation with respect to $\sigma$. 
By evaluating Eqs.(\ref{P1}, \ref{fNL1}) with field fluctuations at $t=t_*$ 
such that $\mathcal{P}_{\delta\sigma_*}=(H_*/2\pi)^2$ and then using Eq.(\ref{n'n''}), we obtain
\begin{equation} \label{P2}
\mathcal{P}_{\zeta}=\left[\frac{1}{6}\frac{\Gamma'(\sigma_{\rm{reh}})}{\Gamma(\sigma_{\rm{reh}})}\frac{\partial\sigma_{\rm{reh}}}{\partial\sigma_*}\right]^2\left(\frac{H_*}{2\pi}\right)^2,
\end{equation}
\begin{equation} \label{fNL2}
f_{\rm{NL}}=5\left[1-\frac{\Gamma(\sigma_{\rm{reh}})\Gamma''(\sigma_{\rm{reh}})}{\Gamma'(\sigma_{\rm{reh}})^2}-\frac{\Gamma(\sigma_{\rm{reh}})}{\Gamma'(\sigma_{\rm{reh}})}\left(\frac{\partial\sigma_{\rm{reh}}}{\partial\sigma_*}\right)^{-2}\frac{\partial^2\sigma_{\rm{reh}}}{\partial\sigma_*^2}\right].
\end{equation}
These equations fully account for the evolution of the modulus between horizon crossing and reheating. 
If the modulus rolls after the perturbation modes exit the horizon, then the rolling will
affect $\mathcal{P}_{\zeta}$ and $f_{\rm{NL}}$ due to the factors of $\partial\sigma_{\rm{reh}}/\partial\sigma_*$ and $\partial^2\sigma_{\rm{reh}}/\partial\sigma_*^2$ in Eqs.(\ref{P2}, \ref{fNL2}). These factors can be written in terms of the modulus potential by solving the modulus equation of motion (EOM): 
\begin{equation} \label{EOM}
\ddot\sigma+3H\dot\sigma+V'(\sigma)=0.
\end{equation}

To solve the modulus EOM analytically, we rewrite it in the form 
\begin{equation} \label{cEOM}
c(t)H\dot\sigma+V'(\sigma)=0,
\end{equation}
with 
\begin{equation} \label{c}
 c(t) = \begin{dcases}
        3, \phantom{aaai} t\leq t_{\rm{end}}\\
      \frac{9}{2}, \phantom{aaa} t_{\rm{end}}<t<t_{\rm{reh}}.
        \end{dcases}
\end{equation}
Equation (\ref{cEOM}) with $c(t)$ given by Eq.(\ref{c}) is an accurate approximation to the exact equation of motion Eq.(\ref{EOM}), provided that
\begin{equation} \label{light}
\left|\frac{V''(\sigma)}{c(t)H^2}\right|\ll1
\end{equation}
before reheating. We will call this inequality the lightness
condition.  In a universe with a constant equation of state parameter $w = p /
\rho$, Eq.(\ref{cEOM}) with $c=3+\frac{3}{2}(w+1)$ 
is an attractor solution for a light modulus~\cite{chiba09,
 kawasaki11}. The inflationary ($w\simeq-1$) and inflaton-oscillation
($w=0$) eras give Eq.(\ref{c}). 
For the rest of this section, we will assume that Eq.(\ref{light}) holds, and we will use Eq.(\ref{cEOM}) to evaluate $\sigma(t)$.  In sections 3 and 4, we will focus on models that satisfy Eq.(\ref{light}), but 
we will also show in section 3 that a marginal violation of 
Eq.(\ref{light}) enhances the impact of the modulus dynamics on the curvature perturbations, as the modulus starts to fast-roll prior to reheating in such cases. The modulated reheating mechanism may work even if the modulus starts to oscillate prior to reheating, but we assume throughout this paper that the inflaton decays before the modulus reaches its potential minimum.

Within the sudden-decay framework, the Universe is effectively dominated by matter until $t_{\rm{reh}}$, and so we can use $c=9/2$ for all times between the end of inflation and reheating. Equipped with Eq.(\ref{c}), we then take the following integral of Eq.(\ref{cEOM}): 
\begin{equation} \label{gint}
\int^{\sigma_{\rm{reh}}}_{\sigma_*}\frac{d\sigma}{V'(\sigma)}=-\int^{\sigma_{\rm{end}}}_{\sigma_*}\frac{d\sigma}{3H\dot\sigma}-\int^{\sigma_{\rm{reh}}}_{\sigma_{\rm{end}}}\frac{2d\sigma}{9H\dot\sigma}=-\frac{1}{3}\int^{t_{\rm{end}}}_{t_*}\frac{dt}{H}-\frac{2}{27}\frac{\beta^2}{\Gamma(\sigma_{\rm{reh}})^2}+\frac{2}{27}\frac{1}{H_{\rm{end}}^2},
\end{equation}
where we have assumed $V'(\sigma)\ne0$. After differentiating both sides with respect to $\sigma_*$, we obtain
\begin{equation} \label{dsds}
\frac{\partial\sigma_{\rm{reh}}}{\partial\sigma_*}=\frac{1}{1-X(\sigma_{\rm{reh}})}\frac{V'(\sigma_{\rm{reh}})}{V'(\sigma_*)},
\end{equation}
where
\begin{equation} \label{V'V'X}
X(\sigma_{\rm{reh}})\equiv\frac{4\beta^2}{27}\frac{\Gamma'(\sigma_{\rm{reh}})V'(\sigma_{\rm{reh}})}{\Gamma(\sigma_{\rm{reh}})^3}.
\end{equation}
Equations (\ref{cEOM}, \ref{c}) imply that $X(\sigma_{\rm{reh}})$ is directly related to the rate at which the decay rate is changing at $t=t_{\rm{reh}}$:
\begin{equation} \label{XGamma}
X(\sigma_{\rm{reh}})=-\frac{2\beta^2}{3}\frac{H(t)\dot\Gamma(\sigma)}{\Gamma(\sigma)^3}\bigg|_{t_{\rm{reh}}}=-\frac{2\beta}{3}\frac{\dot\Gamma(\sigma_{\rm{reh}})}{\Gamma(\sigma_{\rm{reh}})^2},
\end{equation}
where we have used $\Gamma(\sigma_{\rm{reh}})/H(t_{\rm{reh}})=\beta$ to obtain the final expression. 

It is clear from Eq.(\ref{dsds}) that $|X(\sigma_{\rm{reh}})|$ provides a measure of the importance of the modulus's evolution; if $|X(\sigma_{\rm{reh}})|\gtrsim1$, then $\partial\sigma_{\rm{reh}}/\partial\sigma_*$ can significantly differ
from unity.  It follows from Eqs.(\ref{P2}, \ref{fNL2}) that
$\mathcal{P}_{\zeta}$ and $f_{\rm{NL}}$ are significantly affected by
the modulus's evolution if $|X(\sigma_{\rm{reh}})|\gtrsim1$. We present a model with $X(\sigma_{\rm{reh}}) < -1$ in section 3.1, and we show
that the rolling of the modulus suppresses the amplitude of the power spectrum while enhancing the amplitude of the bispectrum. We note that $X(\sigma_{\rm{reh}})$ can only take positive values if the inflaton decay rate decreases with time [see Eq.(\ref{XGamma})]. Since $\Gamma$ is initially less than $H$, the decay condition $\Gamma(\sigma_{\rm{reh}})/H(t_{\rm{reh}})=\beta$ can be satisfied only if $\Gamma$ is decreasing slower than $H$ at reheating.  Demanding that $\beta \dot{H}(t_{\mathrm{reh}}) \leq\dot{\Gamma}(\sigma_{\mathrm{reh}})$ restricts the range of possible $X(\sigma_{\mathrm{reh}})$ values to
\begin{equation}
 X(\sigma_{\mathrm{reh}}) \lesssim 1.
\end{equation}
Therefore, large values of $|X(\sigma_{\rm{reh}})|$ are only possible if the inflaton decay rate is increasing at the time of reheating.

Since $|X(\sigma_{\rm{reh}})|$ takes non-negligible values only when
$\Gamma(\sigma)$ varies during reheating, we must consider the
implications of an evolving decay rate.   To ensure that $\Gamma$ can
still be interpreted as the decay rate of a massive particle, we will
assume that the inflaton mass $m_\phi$ is always larger than
$|\dot{\Gamma}/\Gamma|$, so that $\Gamma$ is nearly constant during each
oscillation of the inflaton field.  Even with this assumption, however,
a time-varying $\Gamma$ does not necessarily lead to exponential decay
of the form $\rho_\phi \propto e^{-\Gamma t}$. 
The assumption that the inflaton decays instantly is therefore not
necessarily warranted when $|X(\sigma_{\rm{reh}})|\gtrsim1$.  In
particular, it is clear that Eq.(\ref{dsds}) must not apply when
$X(\sigma_{\rm{reh}})=1$.  If the instant decay of the inflaton is an
inaccurate approximation, we must numerically solve the modulus field's EOM and the evolution equations for the inflaton and radiation
energy densities [provided in Eq.(\ref{cont2})].  In the following sections, we
use these numerical solutions to test the validity of the sudden-decay
approximation. 
We will see that $|X(\sigma_{\rm{reh}})| \gtrsim 1 $ does not necessarily invalidate the
sudden-decay approximation.
Even in cases where the sudden-decay picture
is invalid, the analytic expressions for $\mathcal{P}_{\zeta}$ and $f_{\rm{NL}}$ given by Eqs.(\ref{P3}, \ref{fNL3})
will be useful for understanding the overall behaviour of the resulting density perturbations (as we will see in section 3.2).

With Eq.(\ref{dsds}) and its derivative with respect to $\sigma_*$, we can evaluate Eqs.(\ref{P2}, \ref{fNL2}) to obtain
\begin{equation} \label{P3}
\mathcal{P}_{\zeta}=\left[\frac{1}{6}\frac{\Gamma'(\sigma_{\rm{reh}})}{\Gamma(\sigma_{\rm{reh}})}\frac{H_*}{2\pi}\right]^2\left[\frac{1}{1-X(\sigma_{\rm{reh}})}\frac{V'(\sigma_{\rm{reh}})}{V'(\sigma_*)}\right]^2,
\end{equation}
\begin{align} \label{fNL3}
f_{\rm{NL}}=5\Bigg\{&1-\frac{\Gamma(\sigma_{\rm{reh}})\Gamma''(\sigma_{\rm{reh}})}{\Gamma'(\sigma_{\rm{reh}})^2}+\frac{X(\sigma_{\rm{reh}})}{1-X(\sigma_{\rm{reh}})}\left[3-\frac{\Gamma(\sigma_{\rm{reh}})\Gamma''(\sigma_{\rm{reh}})}{\Gamma'(\sigma_{\rm{reh}})^2}\right] \nonumber\\
&-\frac{4\beta^2}{27}\left[\frac{1}{1-X(\sigma_{\rm{reh}})}\frac{V''(\sigma_{\rm{reh}})}{\Gamma(\sigma_{\rm{reh}})^2}+\frac{V''(\sigma_*)}{\Gamma(\sigma_{\rm{reh}})^2}\right]
-\frac{4\beta^2}{27}\frac{1}{X(\sigma_{\rm{reh}})}\frac{V''(\sigma_{\rm{reh}})-V''(\sigma_*)}{\Gamma(\sigma_{\rm{reh}})^2}\Bigg\}.
\end{align}
The approximate equation of motion, Eq.(\ref{cEOM}), can provide  $\sigma_{\rm{reh}}$ as a function of $\sigma_*$, which can then be used to express $\mathcal{P}_{\zeta}$ and $f_{\rm{NL}}$ as functions of $\sigma_*$.

\subsection{Implications of the Analytic Expressions}
Before we investigate specific reheating scenarios in section 3, we present some general predictions of Eqs.(\ref{P3}, \ref{fNL3}). We will see that the models for which modulus-rolling effects are important can be divided into two classes: (1) models with rapidly evolving inflaton decay rates during reheating, and (2) models in which the effective mass of the modulus changes between inflation and reheating.

We first point out that we can obtain the expressions derived previously\cite{zaldarriaga04, bartolo04, vernizzi04, ichikawa08} by assuming that the modulus does not evolve in Eqs.(\ref{P2}, \ref{fNL2}). Indeed, by setting $\sigma_{\rm{reh}}=\sigma_*$ in these equations, we find: 
\begin{equation} \label{P3no}
\mathcal{P}_{\zeta}=\left(\frac{1}{6}\frac{\Gamma'}{\Gamma}\frac{H_*}{2\pi}\right)^2,
\end{equation}
\begin{equation} \label{fNL3no}
f_{\rm{NL}}=5\left[1-\frac{\Gamma\Gamma''}{(\Gamma')^2}\right].
\end{equation}
Refs.\cite{suyama08, elliston13} go beyond Eqs.(\ref{P3no}, \ref{fNL3no}) by considering the modulus's evolution during inflation, but not after inflation. Since $X(\sigma_{\rm{reh}})$ is proportional to $\dot{\Gamma}$ at reheating, the expressions they find and our results are consistent if we set $\sigma_{\rm{end}}=\sigma_{\rm{reh}}$ and $X(\sigma_{\rm{reh}})=0$ in Eq.(\ref{dsds}).\footnotemark\footnotetext{Ref.\cite{suyama08} allows for the possibility of having non-negligible self-interactions of $\sigma$ and the possibility of having multiple decay channels that depend on more than one modulus fields, and they also keep quantum fluctuations of the inflaton as a source of density fluctuations. Therefore, to be precise, our formalism with $\dot{\sigma}=0$ after inflation agrees with the expressions in Ref.\cite{suyama08} if one considers only one modulus with no self-interactions while also neglecting the contribution to the perturbations from the inflaton field.} 

As we pointed out in the previous section, the size of $|X(\sigma_{\rm{reh}})|$ is an important measure of the effects of the modulus's evolution.  In the limit $|X(\sigma_{\rm{reh}})|\gg1$, Eq.(\ref{fNL3}) reduces to
\begin{multline} \label{Xbig}
\qquad \quad
f_{\rm{NL}}=-10
+\mathcal{O}\left(\frac{1}{X(\sigma_{\rm{reh}})}\right)\times\frac{\Gamma(\sigma_{\rm{reh}})\Gamma''(\sigma_{\rm{reh}})}{\Gamma'(\sigma_{\rm{reh}})^2}
\\ 
+\mathcal{O}\left(\frac{1}{X(\sigma_{\rm{reh}})}, \, 
\frac{V''(\sigma_{\rm{reh}})}{\Gamma(\sigma_{\rm{reh}})^2}
\frac{1}{X(\sigma_{\rm{reh}})^2}, \, 
\frac{V''(\sigma_*)}{\Gamma(\sigma_{\rm{reh}})^2}\right).
\qquad \quad
\end{multline}
We note that $V''(\sigma_*) \sim V''(\sigma_{\mathrm{reh}})$ holds generically for power-law potentials $V(\sigma) \propto \sigma^p$ and thus $V''(\sigma_*)/\Gamma(\sigma_\mathrm{reh})^2\ll1$ as long as the lightness condition Eq.(\ref{light}) applies. 
Unless $|\Gamma(\sigma_{\rm{reh}})\Gamma''(\sigma_{\rm{reh}})/\Gamma'(\sigma_{\rm{reh}})^2|$ is large, we can then make a general prediction that the modulus's rolling modifies $f_{\rm{NL}}$ from Eq.(\ref{fNL3no}) to $f_{\rm{NL}}\simeq-10$ whenever $|X(\sigma_{\rm{reh}})|\gg1$ and $V(\sigma)\propto\sigma^p$, regardless of the form of $\Gamma(\sigma)$. 
In section 3.1, we will present a model in which large values of $|X(\sigma_{\rm{reh}})|$ are realized and $f_{\rm{NL}}$ asymptotes to $-10$ as predicted by Eq.(\ref{Xbig}).

It is important, however, to keep in mind that Eqs.(\ref{P3}, \ref{fNL3}) are based on the assumption that the modulus stays light while the inflaton dominates the energy density of the Universe. As one may expect, the rolling effects become more significant if $V''(\sigma)\gtrsim H^2$ prior to reheating. We will demonstrate the significance of violating the lightness condition by considering such scenarios in section 3.1.  In these instances, the analytic expressions given by Eqs.(\ref{P3}, \ref{fNL3}) do not provide the correct results, and we will see that $f_{\rm{NL}}$ can reach values that are far more negative than $-10$.
Equations (\ref{P3}, \ref{fNL3}) are also invalid if an instantaneous transition to radiation domination inadequately describes the inflaton's decay process. As we mentioned earlier, it is particularly clear that Eqs.(\ref{P3}, \ref{fNL3}) are ill-behaved when $X(\sigma_{\rm{reh}})=1$. The sudden-decay approximation becomes inaccurate if the inflaton decay rate evolves in a non-trivial way, which can happen in some cases with $|X(\sigma_{\rm{reh}})|\gtrsim1$. We present a simple model that realizes a violation of the sudden-decay approximation in section 3.2.

The modulus dynamics become relevant when $|X(\sigma_{\rm{reh}})|$ is large, but $|X(\sigma_{\rm{reh}})|\ll1$ does not imply that the modulus dynamics can be neglected. When $|X(\sigma_{\rm{reh}})|$ is small, $f_{\rm{NL}}$ can still differ significantly from Eq.(\ref{fNL3no}). Assuming that $X(\sigma_{\rm{reh}})|\ll1$, Eq.(\ref{fNL3}) reduces to 
\begin{align}  \label{Xsmall}
f_{\rm{NL}}=&5\left\{1-\frac{\Gamma(\sigma_{\rm{reh}})\Gamma''(\sigma_{\rm{reh}})}{\Gamma'(\sigma_{\rm{reh}})^2}-\frac{4\beta^2}{27}\frac{1}{X(\sigma_{\rm{reh}})}\frac{V''(\sigma_{\rm{reh}})-V''(\sigma_*)}{\Gamma(\sigma_{\rm{reh}})^2}\right\} \nonumber\\
&+\mathcal{O}(X(\sigma_{\rm{reh}}))\times\frac{\Gamma(\sigma_{\rm{reh}})\Gamma''(\sigma_{\rm{reh}})}{\Gamma'(\sigma_{\rm{reh}})^2} +\mathcal{O}\left(X(\sigma_{\rm{reh}}),\frac{V''(\sigma_{\rm{reh}})}{\Gamma(\sigma_{\rm{reh}})^2},\frac{V''(\sigma_*)}{\Gamma(\sigma_{\rm{reh}})^2}\right).
\end{align}
If the effective mass of the modulus is not constant, then the term involving $X(\sigma_{\rm{reh}})^{-1}$ in the first line of Eq.(\ref{Xsmall}) may significantly modify $f_{\rm{NL}}$ from Eq.(\ref{fNL3no}). We devote section 4 to a reheating scenario in which $|X(\sigma_{\rm{reh}})|\ll1$ and yet $f_{\rm{NL}}$ is greatly modified due to the evolution of the modulus. 

For completeness, we also provide expressions for the scalar spectral index and its running, derived assuming $|\dot{H}|\ll H^2$ at horizon exit:
\begin{equation} \label{ns}
n_s-1=\frac{d\ln\mathcal{P}_{\zeta}}{d\ln k}\Big|_{k=aH_*}=2\frac{\dot{H}_*}{H^2_*}+\frac{2}{3}\frac{V''(\sigma_*)}{H^2_*},
\end{equation}
\begin{equation} \label{alpha}
\alpha=\frac{dn_s}{d\ln k}\Big|_{k=aH_*}=2\frac{\ddot{H}_*}{H^3_*}-4\frac{\dot{H}_*^2}{H^4_*}-\frac{4}{3}\frac{\dot{H}_*}{H^2_*}\frac{V''(\sigma_*)}{H^2_*}-\frac{2}{9}\frac{V'(\sigma_*)V'''(\sigma_*)}{H_*^4}.
\end{equation}
One should note that Eqs.(\ref{ns}, \ref{alpha}) are not affected by the
dynamics of the modulus (or the inflaton) after horizon
crossing,\footnotemark\footnotetext{Equations (\ref{ns},
\ref{alpha}) follow directly from the slow-roll of the modulus while
scales exit the horizon, and we note that these expressions are valid
even when the sudden-decay approximation does not apply; see
Refs.\cite{byrnes06, kobayashi13} for a proof of Eq.(\ref{ns}) in
general spectator field models.} and therefore our work does not
introduce any new factors into $n_s$ or $\alpha$. Nevertheless, these
observables must be kept within observational bounds while building
physical models. 
Provided that $H$ is nearly constant during inflation and that $|V''(\sigma_*)|\ll H_*^2$, then all the terms in Eqs.(\ref{ns}, \ref{alpha}) are much smaller than one, except for the last term in Eq.(\ref{alpha}). However, for the power-law potentials that we consider in the following sections ($V(\sigma)\propto\sigma^p$ with a non-negative integer $p$), $|V'(\sigma)V'''(\sigma)|\leq V''(\sigma)^2$, and thus the last term in Eq.(\ref{alpha}) is also constrained to be small when the lightness condition is satisfied.

\section{Models with a Quadratic Modulus Potential}
We now consider specific examples that highlight the scenarios in which
the dynamics of the modulus are important. We demonstrate how large
values of $|X(\sigma_{\rm{reh}})|$ can be realized in this section,
while we consider the limit of small $|X(\sigma_{\rm{reh}})|$ in section
4. In section 3.1, we present a model where
$|X(\sigma_{\rm{reh}})|\gg 1$ and show that $f_{\rm{NL}}$ matches Eq.(\ref{Xbig}). Then we study a different model in section 3.2 in which a large $|X(\sigma_{\rm{reh}})|$ leads to the breakdown of the inflaton sudden-decay approximation. 

In this section, we assume a quadratic function for the modulus potential,
\begin{equation} \label{V}
V(\sigma)=\frac{1}{2}m^2\sigma^2.
\end{equation}
We obtain $\sigma(t)$ by assuming that $m^2\ll H^2$ and solving the approximate modulus EOM [see Eq.(\ref{cEOM}) or Eq.(\ref{gint})]; evaluating this function at $t_{\rm{reh}}$ gives
\begin{equation} \label{s*sdec}
\sigma_*=\sigma_{\rm{reh}}\times  \exp\left[{\left(\frac{\mathcal{N_*}}{3}-\frac{2}{27}\right)\frac{m^2}{H_*^2}+\frac{2\beta^2}{27}\frac{m^2}{\Gamma(\sigma_{\rm{reh}})^2}}\right], 
\end{equation}
where we have assumed for simplicity that the Hubble rate stays constant during inflation at the value we denote by $H_*$. 
The lightness condition ($m^2\ll H^2$) implies that the exponential factor in Eq.(\ref{s*sdec}) is typically of order unity, and consequently $V'(\sigma_{\rm{reh}})/V'(\sigma_*)=\sigma_{\rm{reh}}/\sigma_*$ is of order unity as well. 
It then follows that any significant deviation of $\partial\sigma_{\rm{reh}}/\partial\sigma_*$ in Eq.(\ref{dsds}) from unity comes from $X(\sigma_{\rm{reh}})\neq0$. 

We now take a power-law function for the inflaton decay rate of the form
\begin{equation} \label{Gamma}
\Gamma(\sigma)=\mu^{1-n}(\sigma-\sigma_0)^n,
\end{equation}
for some nonzero integer $n$ and mass scales $\mu$ and $\sigma_{0}$. The shift $\sigma_0$ denotes a misalignment of the $\sigma$ values that minimize $V(\sigma)$ and $\Gamma(\sigma)$, which can arise from an inflaton decay channel that is independent of $\sigma$, for instance. We can see from Eq.(\ref{fNL3no}) that, if the rolling of the modulus is neglected, one does not find significant non-Gaussianity with this type of inflaton decay rate,\footnotemark\footnotetext{We made this choice of $\Gamma(\sigma)$ in particular to show that the modulus rolling can enhance non-Gaussianity from $|f_{\rm{NL}}|\leq5$ to measurably large values. We point out that other choices of $\Gamma(\sigma)$ can lead to large non-Gaussianity without including the effects of modulus rolling\cite{zaldarriaga04, ichikawa08}.}  
\begin{equation}
|f_{\rm{NL}}|=\left|5\left(1-\frac{n-1}{n}\right)\right|\leq5.
\end{equation}
Similarly, the power spectrum without considering the modulus's rolling follows from Eq.(\ref{P3no}):
\begin{equation} \label{Pinvno}
\mathcal{P}_{\zeta}(\sigma_*) =\left[\frac{n}{6(\sigma_*-\sigma_0)}\frac{H_*}{2\pi}\right]^2. 
\end{equation}
With our formalism, the power spectrum and $f_{\rm{NL}}$ including the modulus-rolling effects can be derived from Eqs.(\ref{P3}, \ref{fNL3}):
\begin{equation} \label{P4}
\mathcal{P}_{\zeta}=\left[\frac{n}{6(\sigma_{\rm{reh}}-\sigma_0)}\frac{H_*}{2\pi}\right]^2\left[\frac{1}{1-X(\sigma_{\rm{reh}})}\right]^2\exp\left[{-2\left(\frac{\mathcal{N_*}}{3}-\frac{2}{27}\right)\frac{m^2}{H_*^2}-\frac{4\beta^2}{27}\frac{m^2}{\Gamma(\sigma_{\rm{reh}})^2}}\right],
\end{equation}
\begin{equation} \label{fNL4}
f_{\rm{NL}}=5\left\{1-\frac{n-1}{n}+\frac{2n+1}{n}\frac{X(\sigma_{\rm{reh}})}{1-X(\sigma_{\rm{reh}})}-\left[\frac{1}{1-X(\sigma_{\rm{reh}})}+1\right]\frac{4\beta^2}{27}\frac{m^2}{\Gamma(\sigma_{\rm{reh}})^2}\right\},
\end{equation}
with
\begin{equation} \label{X}
X(\sigma_{\rm{reh}})=\frac{4n\beta^2}{27}\frac{m^2}{\Gamma(\sigma_{\rm{reh}})^2}\frac{\sigma_{\rm{reh}}}{\sigma_{\rm{reh}}-\sigma_0}.
\end{equation}
We can then obtain $\mathcal{P}_{\zeta}$ and $f_{\rm{NL}}$ as functions of $\sigma_*$ by using Eq.(\ref{s*sdec}). 
We have assumed that the modulus is light in deriving Eqs.(\ref{P4}, \ref{fNL4}), which implies $m^2\ll\Gamma(\sigma_{\rm{reh}})^2$. A large $|X(\sigma_{\rm{reh}})|$ is therefore realized if $|\sigma_{\rm{reh}}-\sigma_0|\ll|\sigma_{\rm{reh}}|$.  As we discussed after Eq.(\ref{s*sdec}), $\sigma_{\rm{reh}}\simeq\sigma_*$, so the modulus's evolution will generally be important for values of $\sigma_*$ near $\sigma_0$.

We perform model-specific investigations in the following sections to demonstrate that $|X(\sigma_{\rm{reh}})|\gg1$ is possible with a quadratic modulus potential and the inflaton decay rate given in Eq.(\ref{Gamma}). We will also compare our analytic work to numerical solutions that do not assume sudden decay of the inflaton. We obtain these solutions by modeling the inflaton energy loss, due to its decay into radiation, as
 \begin{equation} \nonumber 
\dot\rho_{\phi}=-3H\rho_{\phi}-\Gamma(\sigma)\rho_{\phi};
\end{equation} 
\begin{equation}\nonumber 
\dot\rho_{\gamma}=-4H\rho_{\gamma}+\Gamma(\sigma)\rho_{\phi};
\end{equation} 
\begin{equation}\label{cont2}
3 M_{\rm{PL}}^2 H^2=\rho_{\phi}+\rho_\gamma, 
\end{equation}
where $M_{\rm{PL}}$ is the reduced Planck mass. We omit the contribution of $\rho_{\sigma}$ to the Hubble rate because the modulus energy density stays subdominant to the total energy density in our models.  We numerically solve this set of coupled evolution equations together with the modulus EOM, Eq.(\ref{EOM}), to obtain the e-folding number at the final hypersurface of uniform energy density at $t_f$. We first solve the modulus EOM between horizon crossing and the end of inflation at $t_{\rm{end}}$ by setting the initial value $\sigma_*$, the energy scale of inflation $H_*$, and the duration of inflation after horizon exit $\mathcal{N}_*$.  For simplicity, we take the Hubble parameter to be constant during inflation [as we did to obtain Eq.(\ref{s*sdec})].  Once we obtain $\sigma(t_{\rm{end}})$, we use it as the initial condition to solve the modulus EOM, along with Eq.(\ref{cont2}), from the end of inflation (when $\rho_{\phi}=3 M_{\rm{PL}}^2 H_{*}^2$ and $\rho_{\gamma}=0$) until the total energy density reaches some fixed energy density that defines $t_f$ in the radiation-dominated Universe. We test the accuracy of the sudden-decay approximation by comparing this numerical solution to the analytic approximation given by Eqs.(\ref{P3}, \ref{fNL3}).

Equation (\ref{cont2}) implicitly assumes that the inflaton's oscillation rate, $m_{\phi}$, is much larger than both $H$ and $\Gamma$. While this condition is not satisfied immediately at the end of inflation, it will generally be satisfied as the Hubble rate decreases after inflation. In all modulated reheating scenarios, the Universe is effectively matter dominated prior to reheating, which implies that $m_{\phi}\gg H$ when the inflaton decay rate is important (when $H \simeq \Gamma$).

\subsection{A Monotonically Increasing Decay Rate}
\begin{figure}[t]
\centering
\includegraphics[width=3.1in]{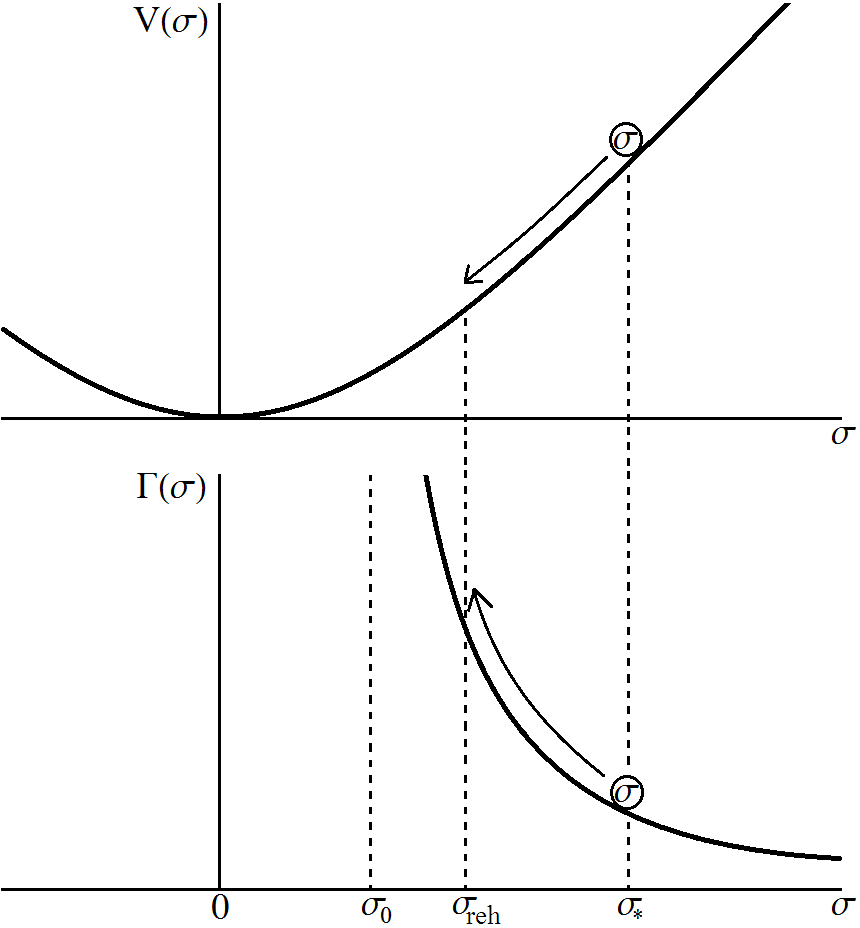}
\caption{Illustration of the quadratic modulus potential $V(\sigma)$ and the inflaton decay rate $\Gamma(\sigma)$ for the model in section 3.1. The inflaton decay rate is inversely proportional to $\sigma-\sigma_0$, with $\sigma_0 > 0$.  The modulus initially has a field value larger than $\sigma_0$, and it rolls toward $\sigma_0$. Reheating happens when the decay rate surpasses the Hubble rate.}
\label{fig:inv}
\end{figure}

\FloatBarrier
In this section, we investigate a simple model in which large values of $|X(\sigma_{\rm{reh}})|$ are realized, and thus the power spectrum and the non-linearity parameter in Eqs.(\ref{P4}, \ref{fNL4})  receive significant contributions from the modulus dynamics. We take the inflaton decay rate to be inversely related to the modulus field, given by Eq.(\ref{Gamma}) with $ n=-1$:
\begin{equation} \label{gammainv}
\Gamma(\sigma)=\mu^2(\sigma-\sigma_0)^{-1},
\end{equation}
with $\sigma_*>\sigma_0>0$ (see Figure \ref{fig:inv} for an
illustration of this model).\footnotemark\footnotetext{The main utility
of this model is its simplicity, but it may be possible to realize such
a decay rate with a modulus-dependent inflaton mass,
$m_{\phi}=m_{\phi}(\sigma)$. 
An interaction term with a light scalar~$\chi$ such as  $\mathcal{L}\ni
\lambda\phi\chi^2$ sources a decay rate $\Gamma\sim \lambda^2 /
m_{\phi}(\sigma)$, which may give rise to a decay rate proportional to
$\sigma^{-1}$.} 
Since the modulus field rolls toward $\sigma_0$, the decay rate is a monotonically increasing function of time. 

With this choice of $\Gamma(\sigma)$, the relationship between $\sigma_*$ and $\sigma_{\rm{reh}}$ in Eq.(\ref{s*sdec}) can be approximated as 
\begin{equation} \label{mess0}
\sigma_{\rm{reh}}=\sigma_*\times q^{-1}  \left[1-\frac{2\beta^2}{27}\frac{m^2}{\Gamma(\sigma_{\rm{reh}})^2}\right]
\end{equation}
where $q\equiv \exp\left[{\left(\frac{\mathcal{N_*}}{3}-\frac{2}{27}\right)\frac{m^2}{H_*^2}}\right]$ and we have neglected terms of order $m^4/\Gamma(\sigma_{\rm{reh}})^4$. This equation can be solved analytically as
\begin{equation} \label{mess}
\sigma_{\rm{reh}}=\sigma_0+\frac{\sqrt{\frac{8\beta^2}{27}\frac{m^2}{\mu^4}q^{-1}\sigma_*\left(q^{-1}\sigma_*-\sigma_0\right)+1}-1}{\frac{4\beta^2}{27}\frac{m^2}{\mu^4}q^{-1}\sigma_*}.
\end{equation}
By substituting Eq.(\ref{mess}) into Eq.(\ref{P4}) for $\mathcal{P}_{\zeta}$, we can obtain an analytic expression for the power spectrum as a function of $\sigma_*$.  We can further simplify Eq.(\ref{mess}), and thus the formula for $\mathcal{P}_{\zeta}(\sigma_*)$, in the limit $|X(\sigma_{\rm{reh}})|\gg1$. First, with the decay rate given in Eq.(\ref{gammainv}) we obtain
\begin{equation} \label{Xinv}
X(\sigma_{\rm{reh}})=-\frac{4\beta^2}{27}\frac{m^2}{\mu^4}\sigma_{\rm{reh}}(\sigma_{\rm{reh}}-\sigma_0).
\end{equation}
By construction we have $\sigma_0<\sigma_{\rm{reh}}<q^{-1}\sigma_*$, so 
\begin{equation} 
|X(\sigma_{\rm{reh}})| < \frac{4\beta^2}{27}\frac{m^2}{\mu^4}q^{-1}\sigma_*\left(q^{-1}\sigma_*-\sigma_0\right)
\end{equation}
and hence in the limit $|X(\sigma_{\rm{reh}})|\gg1$, we can reduce Eq.(\ref{mess}) to
\begin{equation} \label{Eq43}
\sigma_{\rm{reh}}=\sigma_0+\sqrt{\frac{\sigma_*-q\sigma_0}{\frac{2\beta^2}{27}\frac{m^2}{\mu^4}\sigma_*}}\left[1+\mathcal{O}\left(\frac{1}{|X(\sigma_{\rm{reh}})|^{1/2}}\right)\right].
\end{equation}
Inserting $\sigma_{\rm{reh}}$ into the power spectrum Eq.(\ref{P4}), we finally obtain
\begin{equation} \label{1/4}
\mathcal{P}_{\zeta}=\frac{1}{4}\left[\frac{1}{6(\sigma_*-q\sigma_0)}\frac{H_*}{2\pi}\right]^2\left[1+\mathcal{O}\left(\frac{1}{|X(\sigma_{\rm{reh}})|^{1/2}}\right)\right].
\end{equation}
We assume the modulus to be light during inflation, $m^2/H_*^2\ll1$, which implies that $q$ is slightly larger than unity. In contrast, $\mathcal{P}_{\zeta}(\sigma_*)$ is given by Eq.(\ref{Pinvno}) with $n=-1$ if the modulus dynamics are neglected. We therefore see that the modulus rolling suppresses the power spectrum by a factor that is slightly smaller than $4$ in the limit of large $|X(\sigma_{\rm{reh}})|$. Recall from the discussion following Eq.(\ref{Xbig}) that we also expect to see $f_{NL}\simeq-10$ in the limit $|X(\sigma_{\rm{reh}})|\gg1$. 

We now adopt the following set of parameters as a definite example for which large values of $|X(\sigma_{\rm{reh}})|$ are realized, and thus the modulus rolling becomes important:
\{$H_*=10^{11}\rm{GeV}$, $\mathcal{N}_*=50$, $m=8.6\times10^{-10}M_{\rm{PL}}$, $\mu=4.5\times10^{-8}M_{\rm{PL}}$, $\sigma_{0}=4.0\times  10^{-4}M_{\rm{PL}}\}$. We treat the remaining parameter $\sigma_*$ as the independent variable. This set is only a representative selection of parameters for which $|X(\sigma_{\rm{reh}})|$ becomes large and it is not the unique set. To obtain this parameter set, we first set a value for $H_*$ such that $\mathcal{P}_\zeta$ from slow-roll inflaton perturbations can be much smaller than $10^{-9}$ without violating current upper bounds on the tensor-to-scalar ratio. For $|X(\sigma_{\rm{reh}})| \gg 1$, setting the power spectrum from modulated reheating near the observed value ($\mathcal{P}_\zeta= 2.2 \times 10^{-9}$\cite{ade13}) roughly fixes $(\sigma_* - q\sigma_0)$, as seen in Eq.(\ref{1/4}).
We then choose values for $\sigma_0$, $m / \mu^2$, and a parameter range for $\sigma_*$ such that $|X(\sigma_{\rm{reh}})| \simeq -20$ and $m^2 / \Gamma (\sigma_{\rm{reh}})^2 \lesssim 1$. The modulus mass $m$ is then chosen such that the modulus is light during inflation ($m^2 \ll H_*^2$), the modulus energy density is negligible compared to the total energy density ($\rho_{\sigma}\ll\rho_{\rm{tot}}$) before reheating, and the classical rolling of the modulus during inflation dominates over the quantum fluctuations, 
\begin{equation} \label{stoc}
\frac{H_*}{2\pi} \ll \frac{|\dot{\sigma}|}{H_*} = \frac{m^2\sigma}{3H_*^2},
\end{equation}
where we used the approximate modulus EOM, Eq.(\ref{cEOM}). Since we
also fix a value for $\mathcal{N}_*$, the wavelength of the perturbation is
uniquely determined by the post-inflation expansion history, which is
set by the remaining parameter $\sigma_*$.
The perturbation scale therefore changes as we vary $\sigma_*$.
 However, the modulus is light during inflation, so the perturbations are nearly scale-invariant, as shown by Eq.(\ref{ns}).

We also set $\beta=1$ in the following calculations so that the time of sudden inflaton decay is defined by $\Gamma(\sigma_{\rm{reh}})/H(t_{\rm{reh}})=1$. Changing $\beta$ mainly results in shifting the e-folding number Eq.(\ref{efold2}) by a constant, which does not affect its slope $\mathcal{N}'(\sigma)$. Although $\beta$ also alters $\sigma_{\rm{reh}}$ by changing $t_{\rm{reh}}$, we note that the observables $\mathcal{P}_{\zeta}$ and $f_{\rm{NL}}$ are rather insensitive to the precise value of $\beta$, as demonstrated explicitly in Eqs.(\ref{1/4}, \ref{Xbig}) for the limit $|X(\sigma_{\rm{reh}})|\gg1$.

Since the modulus's effective mass is constant for a quadratic potential, the lightness condition is satisfied until $H\simeq m$, which occurs $\frac{2}{3}\ln\left(H_*/m\right)= 2.6$ e-foldings after inflation in this model. We require the decay condition $\Gamma(\sigma_{\rm{reh}})/H(t_{\rm{reh}})=1$ to be satisfied before the lightness condition is violated, which implies $\Gamma(\sigma_{\mathrm{reh}}) \gtrsim m $. Applying this inequality to Eq.(\ref{Eq43}), we obtain $\frac{2}{27}\sigma_* \gtrsim \sigma_* - q\sigma_0$ which provides us with a maximum value for the parameter $\sigma_*$: for the parameters we specified, we have $q=1.007$ and thus $\sigma_* \lesssim1.09\sigma_0$. 
The smallest value of $\sigma_*$ we consider must satisfy $\Gamma(\sigma_{\rm{end}}) \ll H_*$ to ensure that the Universe is effectively matter-dominated prior to reheating; even if $\Gamma$ is constant, the sudden-decay approximation introduces an error to the computed power spectrum that scales with the size of $\Gamma(\sigma_{\rm{end}})/H_*$. We restrict our analysis to $\sigma_*\geq1.02\sigma_0$ so that the decay rate at the end of inflation is $\Gamma(\sigma_{\rm{end}}) \lesssim 0.01H_*$.

\begin{figure}[t]
\centering
\includegraphics[trim=1.5cm 1.3cm 0.5cm 6.5cm, width=3.in]{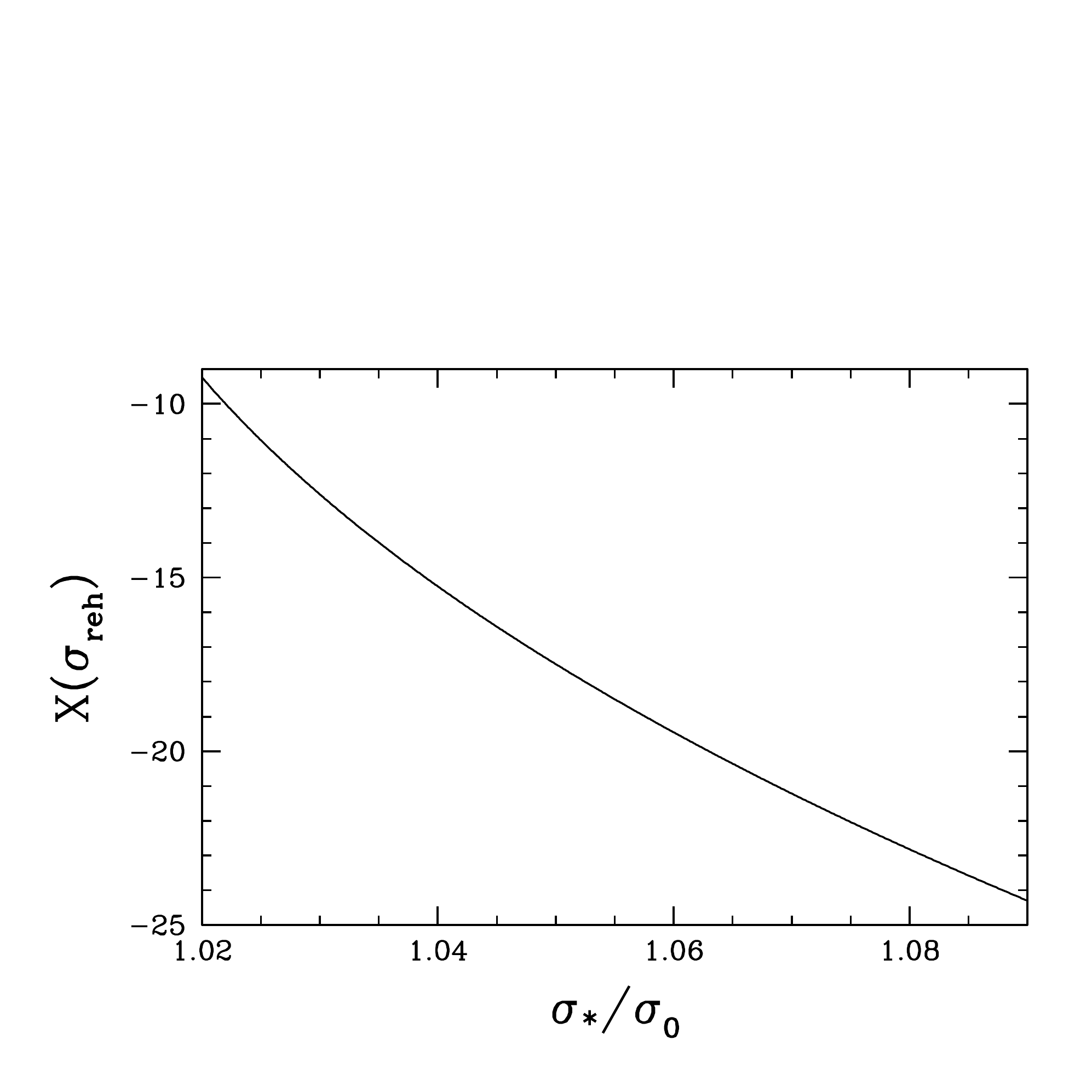}
\includegraphics[trim=1.5cm 1.3cm 0.5cm 6.5cm, width=3.in]{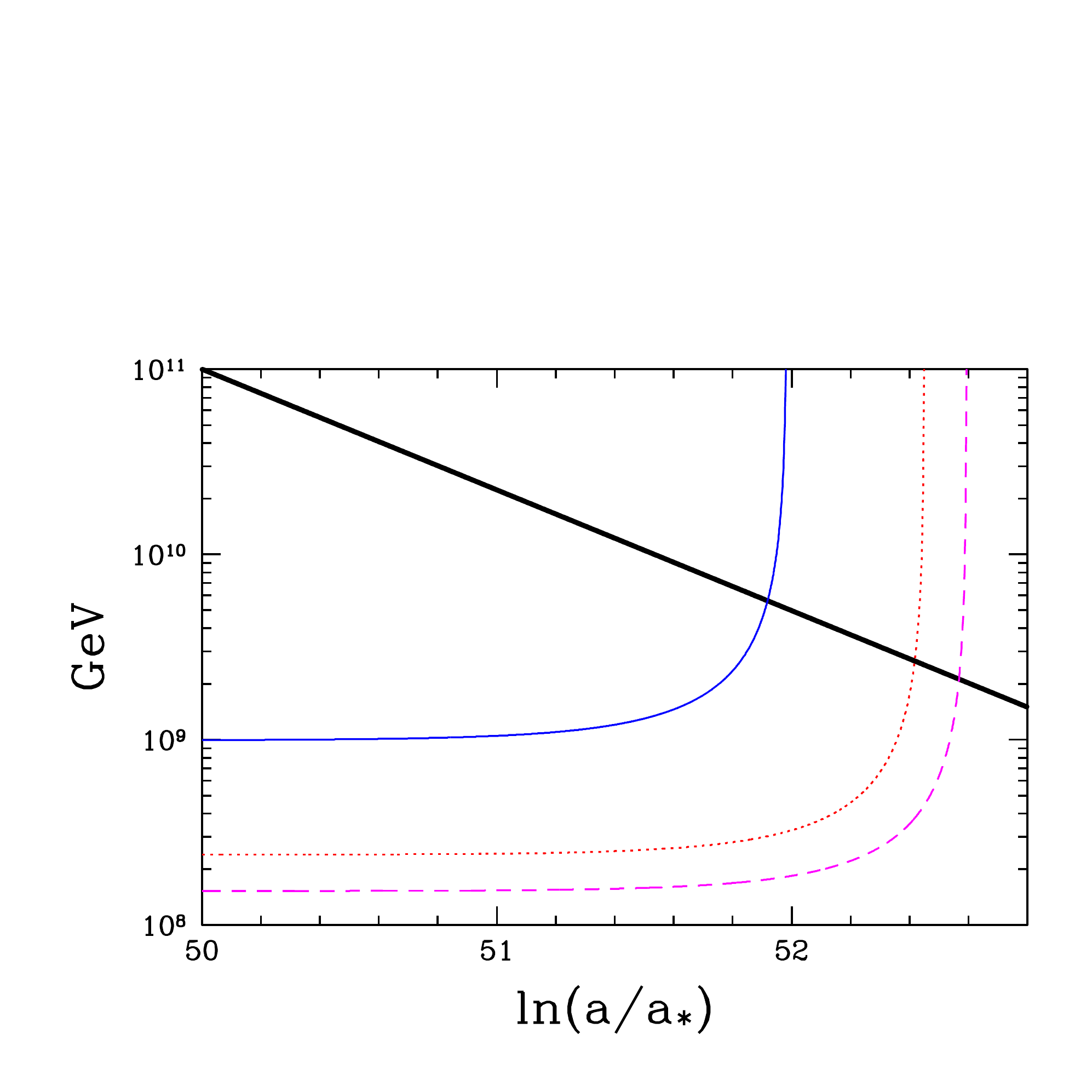}
\caption{Left: plot of $X(\sigma_{\rm{reh}})$, a measure of $\dot\Gamma$
 at reheating, as a function of the modulus value at horizon exit,
 $\sigma_*$. The modulus-rolling effects become significant
 when $|X(\sigma_{\rm{reh}})|\gtrsim1$. Right: plot of the Hubble
 parameter (thick black line) and the decay rate as a function of scale factor
 for three values of $\sigma_*$ near $\sigma_0$: $\sigma_*=1.02\sigma_0$
 (solid blue), $\sigma_*=1.06\sigma_0$ (dotted red), and
 $\sigma_*=1.09\sigma_0$ (dashed magenta). 
Note that the e-folding $\ln (a/a_*) = 50$ corresponds to the end of inflation.}
\label{fig:Xinv}
\end{figure}

We plot $X(\sigma_{\rm{reh}})$ as a function of $\sigma_*$ in the left
panel of Figure \ref{fig:Xinv}; we see that $|X(\sigma_{\rm{reh}})| \gtrsim 10$
for the entire $\sigma_*$ range of interest. 
We can understand the behaviour of $X(\sigma_{\rm{reh}})$ pictorially by
recalling that $X(\sigma_{\rm{reh}})$ is a measure of $\dot{\Gamma}$ at
reheating. We plot $H(t)$ and $\Gamma(\sigma(t))$ for several values of
$\sigma_*$ near $\sigma_0$ in the right panel of Figure
\ref{fig:Xinv}. Under the sudden-decay assumption, the inflaton decays
when the decay condition $\Gamma(\sigma)/H=1$ is satisfied. For the
modulus values starting close to $\sigma_0$, the inflaton decays before
there is significant time for the modulus to roll. As we take $\sigma_*$
further from $\sigma_0$, however, the initial decay rate
$\Gamma(\sigma_*)$ decreases and the modulus has more time to roll
toward $\sigma_0$ before the decay condition is satisfied. The decay
rate when $\Gamma(\sigma)\simeq H$ is clearly rising more rapidly as we
increase the value of $\sigma_*$, and therefore the value of
$|X(\sigma_{\rm{reh}})|$ increases accordingly as we increase
$\sigma_*$.
Such behaviour can also be understood from the equations: 
Eq.(\ref{Xinv}) implies that 
$X(\sigma_{\mathrm{reh}})$ is a negative and monotonically decreasing
function of $\sigma_{\mathrm{reh}}$, as $\sigma_{\mathrm{reh}}> \sigma_0$. Furthermore,
$\partial \sigma_{\mathrm{reh}} / \partial \sigma_* > 0$ for a
negative~$X(\sigma_{\mathrm{reh}})$ [see Eq.(\ref{dsds})]. Hence, the 
monotonic decrease of $X(\sigma_{\mathrm{reh}})$ as a function of
$\sigma_*$ is a generic feature of this model, independent of the choice
of parameters. 
The rapid rise of $\Gamma$ at reheating causes the inflaton decay to 
proceed more rapidly than the usual exponential decay, and we will see that the sudden-decay approximation provides an accurate description of reheating in this scenario.

\begin{figure}[t]
\centering
\includegraphics[trim=1.4cm 1.3cm 0.5cm 6.5cm, width=3.in]{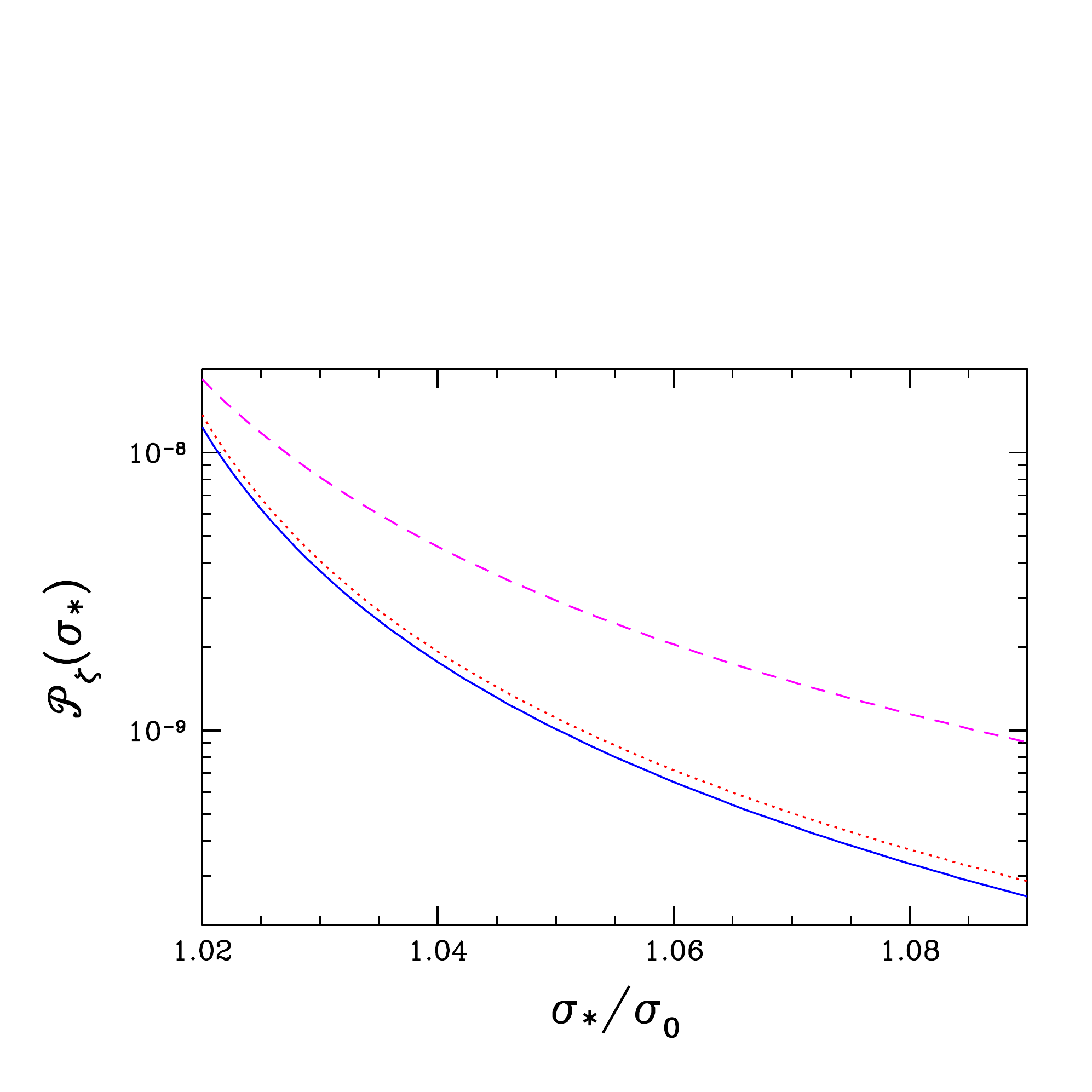}
\includegraphics[trim=1.4cm 1.3cm 0.5cm 6.5cm,  width=3.in]{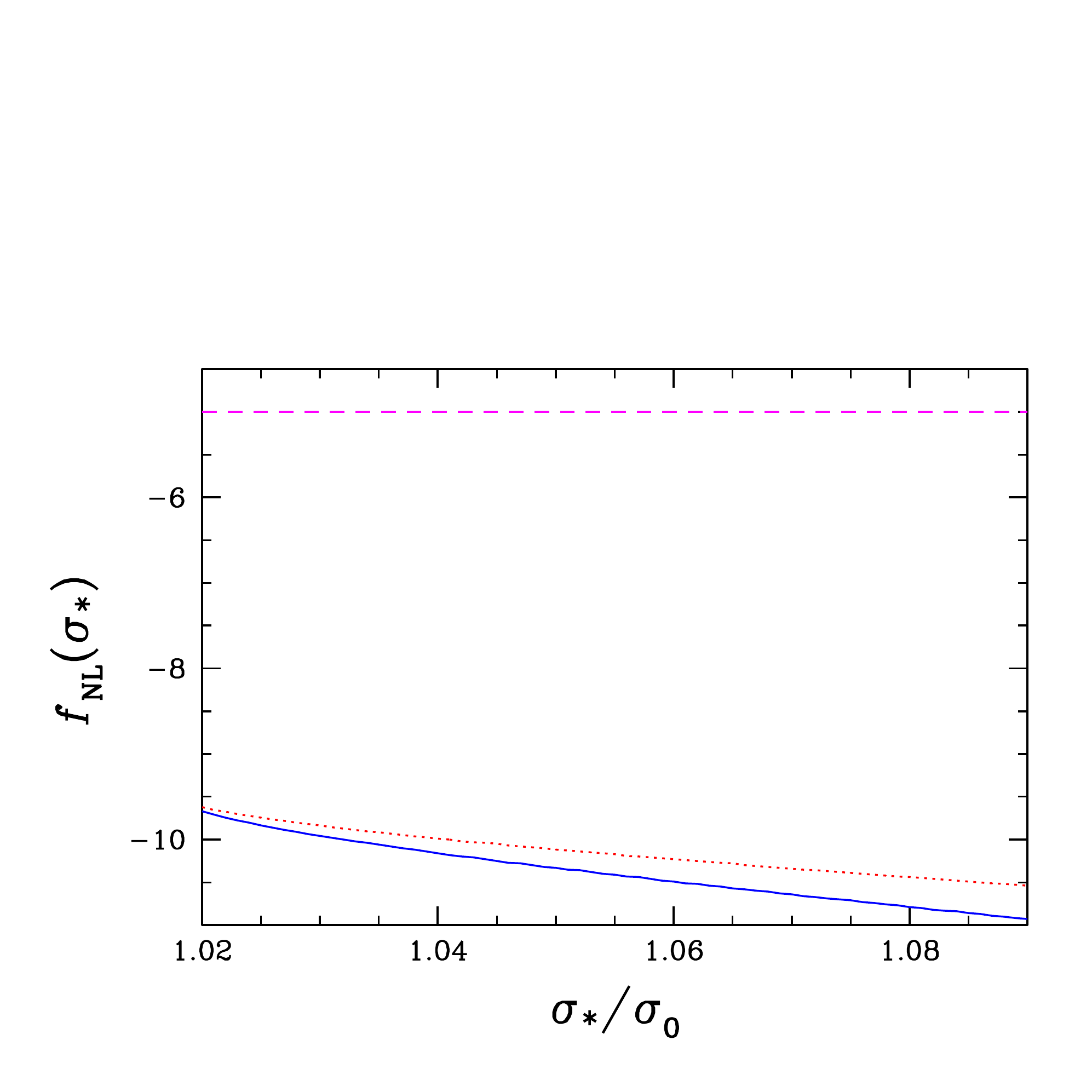}
\caption{The power spectrum (left) and the non-linearity parameter (right) as functions of $\sigma_*$. The solid blue curve comes from the numerical solution. The dotted red curve is our analytic solution derived assuming sudden decay of the inflaton, while the dashed magenta curve assumes sudden decay as well as a fixed modulus value.}
\label{fig:PfNLinv}
\end{figure}

We show the power spectrum and $f_{\rm NL}$ as functions of $\sigma_*$ in Figure \ref{fig:PfNLinv}. The dotted red curves correspond to our analytic solutions Eqs.(\ref{P4}, \ref{fNL4}) with $\sigma_{\rm{reh}}$ given by Eq.(\ref{mess}), which include the modulus rolling effects, while the dashed magenta curves are given by Eqs.(\ref{P3no}, \ref{fNL3no}), which neglect the modulus dynamics. The solid blue line corresponds to the numerical solution obtained by solving the energy transfer equations given in Eq.(\ref{cont2}) with the modulus EOM, Eq.(\ref{EOM}). Figure \ref{fig:PfNLinv} shows that our analytic results match the numerical results well. The power spectrum on the left side of Figure \ref{fig:PfNLinv} is suppressed by a factor of $\lesssim4$ due to the modulus dynamics, as predicted in Eq.(\ref{1/4}). 
The right figure shows that $f_{\rm NL}(\sigma_*)$ is reduced from $-5$ to approximately $-10$, as was predicted in section 2.2. Deviations of $f_{\rm{NL}}(\sigma_*)$ from $-10$ are $\mathcal{O}\left(X(\sigma_{\rm{reh}})^{-1}, m^2/\Gamma(\sigma_{\rm{reh}})^2\right)$, which is consistent with Eq.(\ref{Xbig}). For larger values of $\sigma_*$, Figure \ref{fig:Xinv} shows that $X(\sigma_{\rm{reh}})^{-1}$ is small, and thus the deviation from $f_{\rm{NL}}(\sigma_*)=-10$ can be explained mainly by $m^2/\Gamma(\sigma_{\rm{reh}})^2$, which is allowed to go up to unity. 
 
\begin{figure}[t]
\centering
\includegraphics[trim=1.3cm 1.3cm 0.5cm 6.5cm,  width=3.in]{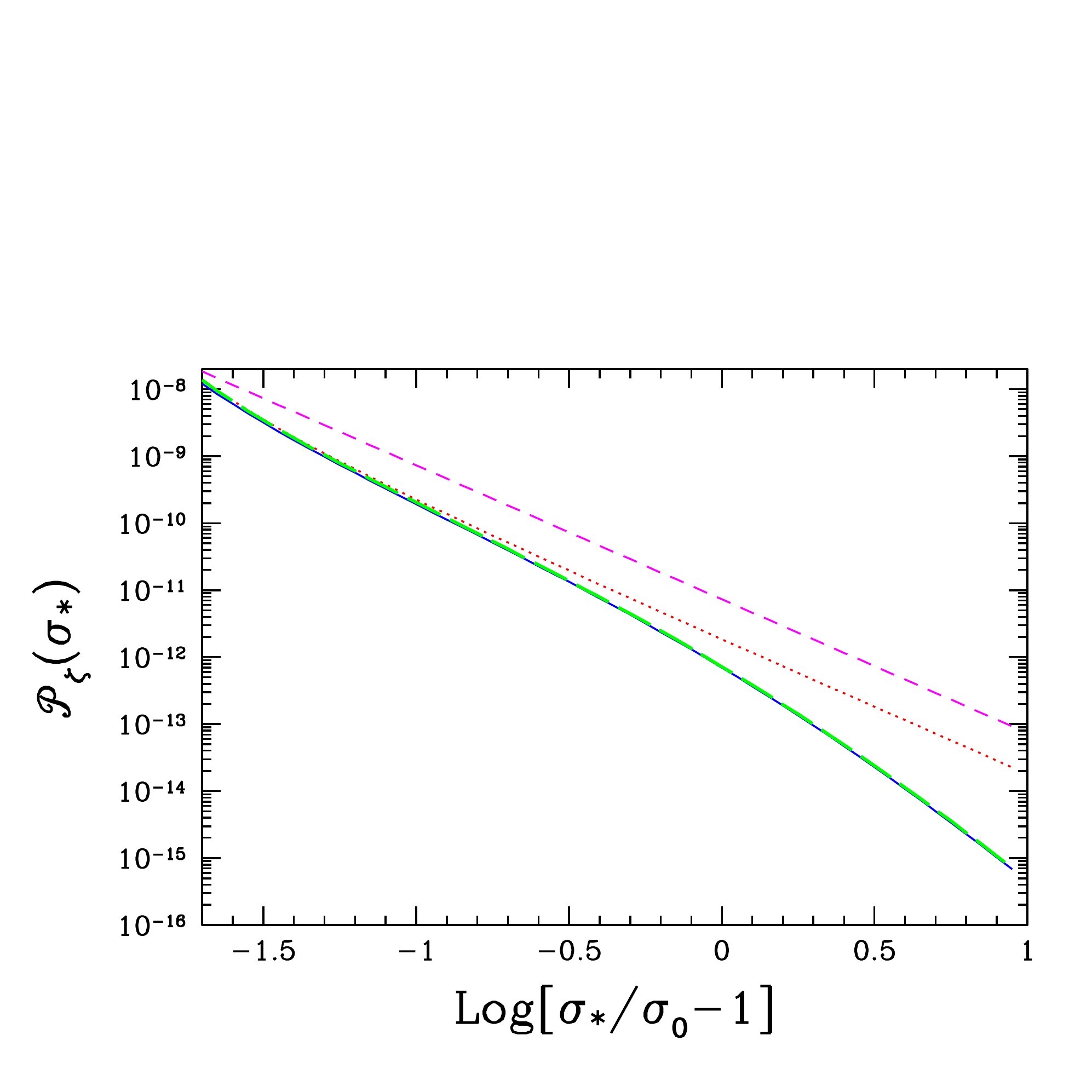}
\includegraphics[trim=1.5cm 1.3cm 0.5cm 6.5cm,  width=3.in]{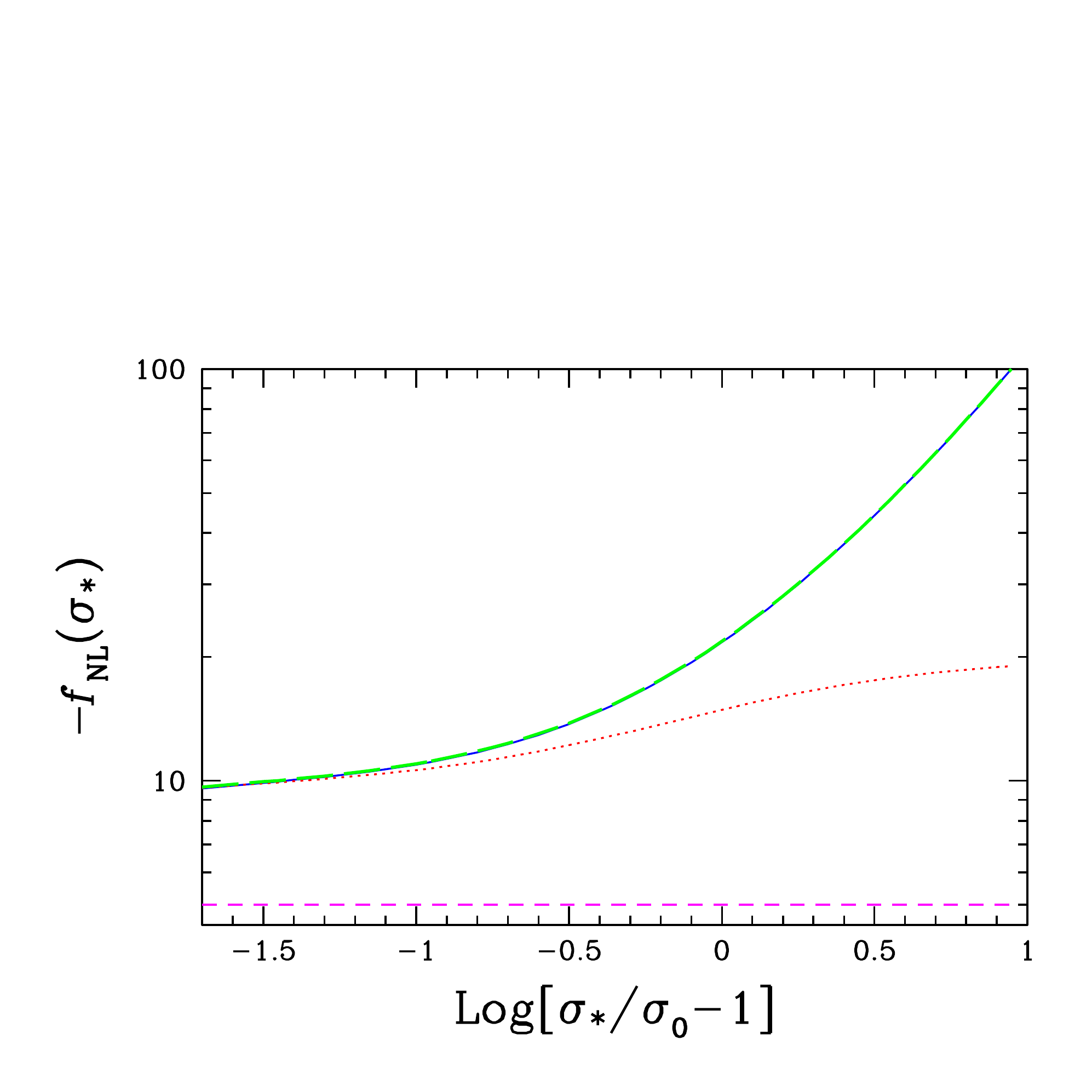}
\caption{The power spectrum (left) and the non-linearity parameter (right) as functions of $\sigma_*$. These plots are extensions of those in Figure \ref{fig:PfNLinv} to larger values of $\sigma_*$. We also add the exact analytic solution (described in the text) in green, long-dashed lines, which lie over the numerical solution (in blue, solid lines).}
\label{fig:PfNLXLinv}
\end{figure}

The analytic formulas do not hold, however, if the lightness condition is violated before the inflaton decays. As one may expect, the modulus dynamics have an even more profound effect if $m^2\gtrsim H^2$ before reheating. In fact, even a mild violation of the lightness condition enhances the non-Gaussianity of the perturbations, with $|f_{\rm{NL}}|$ increasing to values much greater than $10$. In Figure \ref{fig:PfNLXLinv}, we plot the power spectrum and $f_{\rm{NL}}(\sigma_*)$ including values of $\sigma_*$ for which the lightness condition is violated before reheating. We have also plotted green, long-dashed curves that exploit the sudden-decay approximation but do not use the lightness condition; instead of using Eq.(\ref{cEOM}), we use 
\begin{equation}
\sigma(t)=A\times H(t)\cos\left(\frac{2}{3}\frac{m}{H(t)}+\theta\right),
\label{exactsol}
\end{equation}
which is an exact solution to the modulus EOM for a modulus with a quadratic potential in an oscillating-inflaton-dominated universe. The constants $A$ and $\theta$ are found by matching this solution to the slow-roll solution at the end of inflation. Analytic formulas for $\mathcal{P}_{\zeta}(\sigma_*)$ and $f_{\rm{NL}}(\sigma_*)$ are then derived using Eqs.(\ref{P2}, \ref{fNL2}), with the factor of $\partial\sigma_{\rm{reh}}/\partial\sigma_*$ obtained by evaluating Eq.(\ref{exactsol}) when $\Gamma(\sigma)/H=1$. This solution agrees with the numerical solution very well; since it does not assume $m^2\ll H^2$, it illustrates that the breakdown of the lightness condition is responsible for the inaccuracy of our original solutions given in Eqs.(\ref{P4}, \ref{fNL4}). 
Figure \ref{fig:PfNLXLinv} demonstrates that the modulus's evolution can suppress $\mathcal{P}_{\zeta}$ and enhance $f_{\rm{NL}}$ by orders of magnitude if the modulus's mass exceeds the Hubble rate prior to reheating.  

We note that an order-unity modification to $f_{\rm{NL}}$ does not occur over this model's entire parameter space.  Most importantly, if $m^2\ll H^2$, the effects of modulus dynamics are negligible for $\sigma_0=0$ regardless of the values of the other parameters, as Eq.(\ref{Xinv}) gives
\begin{equation}
|X(\sigma_{\rm{reh}})|=\frac{4\beta^2}{27}\frac{m^2}{\Gamma(\sigma_{\rm{reh}})^2}\ll1.
\end{equation}
We can generalize this result to any power-law potential $V(\sigma)\propto\sigma^p$ and decay rate $\Gamma(\sigma)\propto\sigma^n$ with no relative shift of the minimum points (i.e. $\sigma_0=0$). If we write $X(\sigma_{\rm{reh}})$ as 
\begin{equation} \label{Xpower}
X(\sigma_{\rm{reh}})=\frac{4\beta^2}{27}\frac{\Gamma'(\sigma_{\rm{reh}})\sigma_{\rm{reh}}}{\Gamma(\sigma_{\rm{reh}})}\frac{V'(\sigma_{\rm{reh}})}{V''(\sigma_{\rm{reh}})\sigma_{\rm{reh}}}\frac{V''(\sigma_{\rm{reh}})}{\Gamma(\sigma_{\rm{reh}})^2},
\end{equation}
it becomes clear that the size of $|X(\sigma_{\rm{reh}})|$ is usually limited by $V''(\sigma_{\rm{reh}})/\Gamma(\sigma_{\rm{reh}})^2$, which is small for a light modulus;  $|X(\sigma_{\rm{reh}})|$ can only become large if
\begin{equation} \label{gsgvsv}
\left|\frac{\Gamma'(\sigma_{\rm{reh}})\sigma_{\rm{reh}}}{\Gamma(\sigma_{\rm{reh}})}\right|\gg1 \phantom{aa}\text{or}\phantom{aa} \left|\frac{V''(\sigma_{\rm{reh}})\sigma_{\rm{reh}}}{V'(\sigma_{\rm{reh}})}\right|\ll1,
\end{equation}
but neither of these conditions is satisfied if we have $\Gamma(\sigma)\propto\sigma^n$ and $V(\sigma)\propto\sigma^p$, unless $|n| \gg 1$ or $p \simeq 1$.

\subsection{A Quadratic Decay Rate}

\begin{figure}[t]
\centering
\includegraphics[width=3.1in]{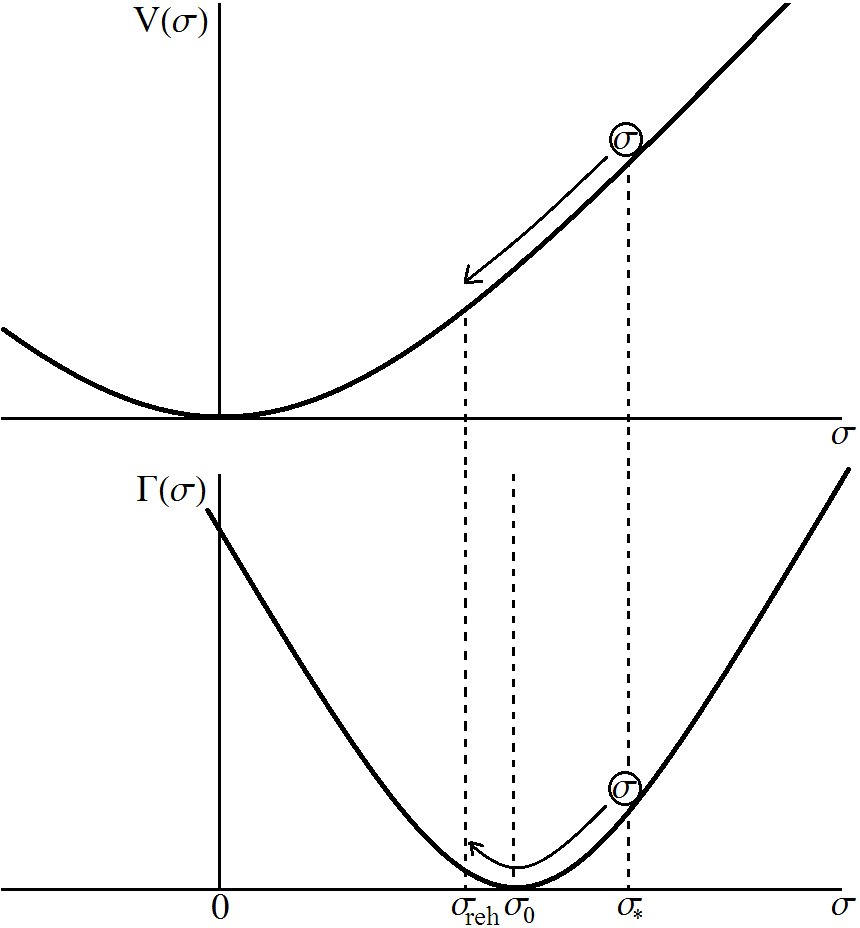}
\caption{Illustration of the quadratic modulus potential $V(\sigma)$ and the inflaton decay rate $\Gamma(\sigma)$ for the model in section 3.2. The inflaton decay rate is a quadratic function of $\sigma$ that has a minimum at $\sigma_0$, with $\sigma_0>0$.  If the modulus initially has a positive field value that is smaller than $\sigma_0$, it rolls away from $\sigma_0$. If $\sigma_*>\sigma_0$ (which is the case illustrated here), the modulus instead rolls towards $\sigma_0$ initially and reheating can happen either before or after the modulus passes through $\sigma_0$.}
\label{fig:quad}
\end{figure}

In this section, we provide another model in which the modulus rolling modifies the observables significantly.  This model also illustrates how the sudden-decay approximation can be inaccurate. We again take the modulus potential to be the quadratic function given in Eq.(\ref{V}), but in this section we take the inflaton decay rate to also be quadratic: 
\begin{equation} \label{gamma2}
\Gamma(\sigma)=\mu^{-1}(\sigma-\sigma_0)^2.
\end{equation}
As in the last section, we are interested in values of $\sigma_*$ near $\sigma_0$ so that $|X(\sigma_{\rm{reh}})|$ is large. In this model, however, the decay rate does not diverge as $\sigma$ approaches $\sigma_0$ (see Figure \ref{fig:quad}). We consider values of $\sigma_*$ both slightly larger and smaller than $\sigma_0$, which allows both positive and negative values of $X(\sigma_{\rm{reh}})$ as seen in Eq.(\ref{X}). In the case of $X(\sigma_{\rm{reh}})=1$, the power spectrum in Eq.(\ref{P3}) diverges, but we will show that this unphysical result is attributable to our use of the sudden-decay approximation in deriving Eq.(\ref{P3}). 

We select the following parameter set using the procedure described in the previous section to ensure that $|X(\sigma_{\rm{reh}})|$ takes large values while satisfying the other constraints on the model: 
\{$H_*=10^{11}\rm{GeV}$, $\mathcal{N}_*=50$, $m=4.6\times10^{-10}M_{\rm{PL}}$, $\mu=6.3\times10^{-1}M_{\rm{PL}}$, $\sigma_{0}=1.3\times10^{-3}M_{\rm{PL}}\}$. 
As in the previous section, we set $\beta = 1$ when evaluating the analytical expressions that employ the sudden-decay approximation.  When the sudden-decay approximation is accurate, the analytical expressions that depend on $\beta$ are rather insensitive to its value.  When the sudden-decay approximation fails, as it does for some values of $\sigma_*$ in this model, the value of $\beta$ does not significantly alter the large discrepancy between the analytical expressions and the numerical results.  Therefore, the precise value of $\beta$ does not affect our analysis of this model.

\begin{figure}[t]
\centering
\includegraphics[trim=1.3cm 1.3cm 0.5cm 6.5cm,  width=3.in]{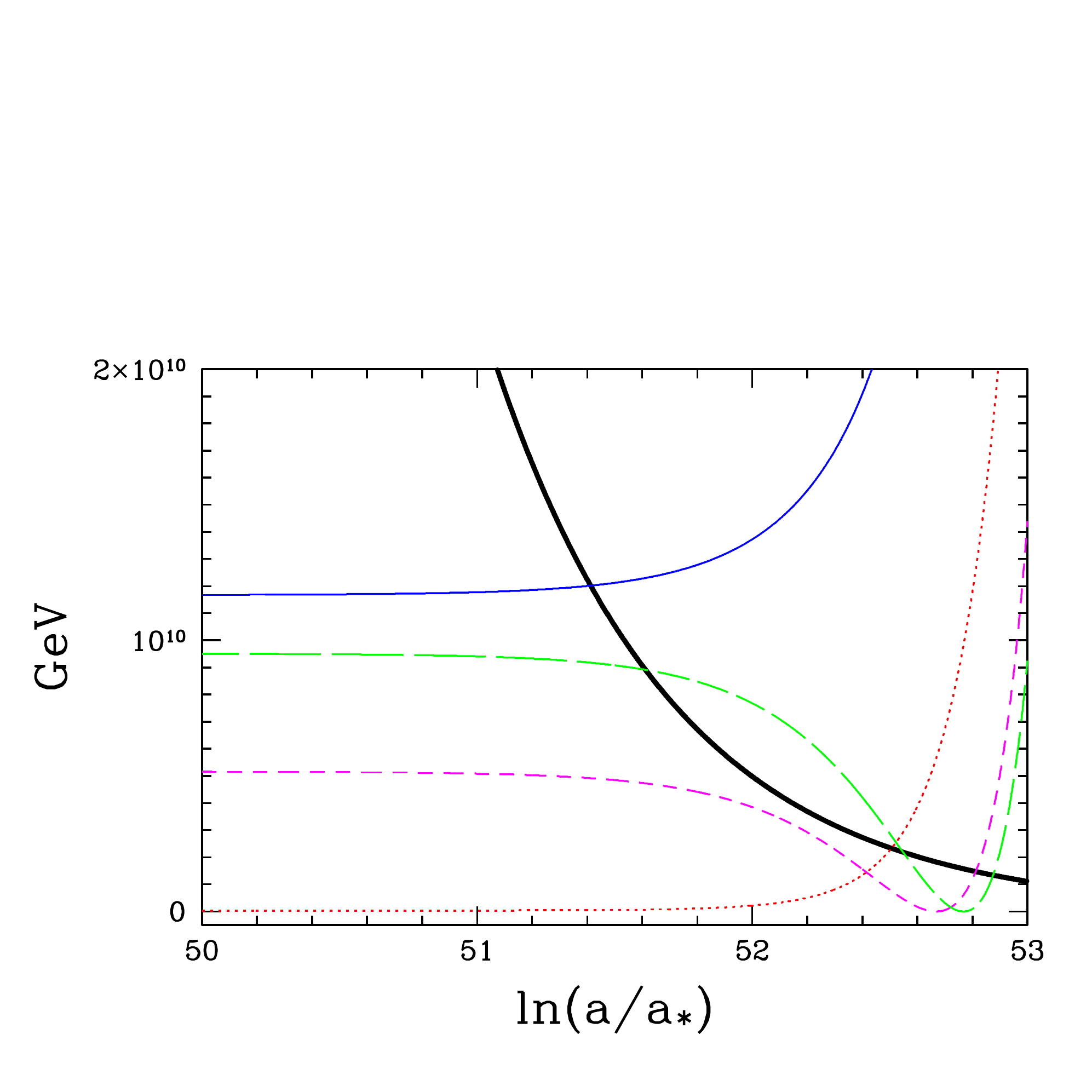}
\includegraphics[trim=1.4cm 1.3cm 0.5cm 6.5cm,  width=3.in]{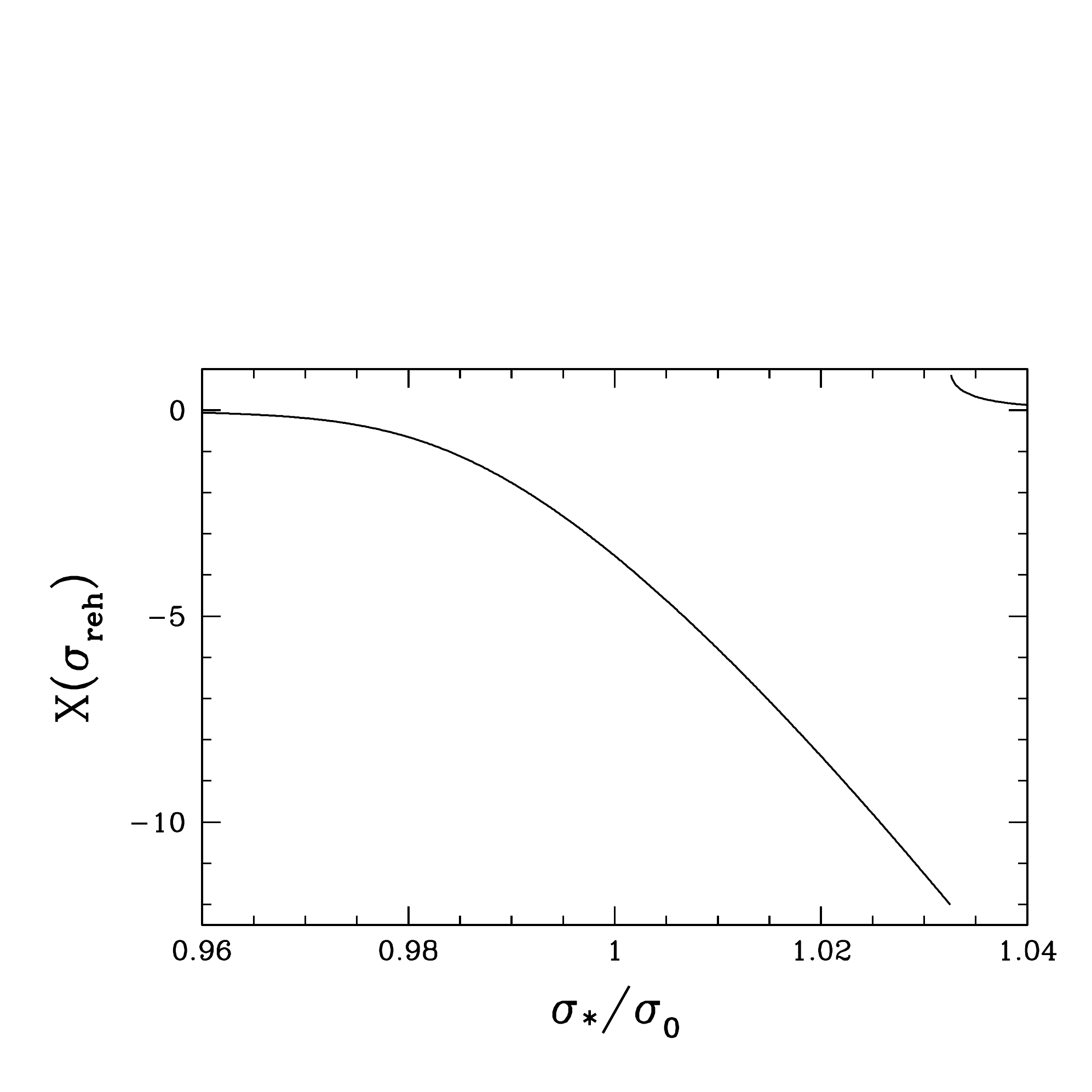}
\caption{Left: plot of the Hubble parameter (thick black line) and the decay rate as a function of scale factor for four values of $\sigma_*$: $0.96\sigma_0$ (solid blue), $1.00\sigma_0$ (dotted red), $1.03\sigma_0$ (dashed magenta), and $1.04\sigma_0$ (long-dashed green). 
Right: Plot of $X(\sigma_{\rm{reh}})$, a measure of $\dot\Gamma$ at reheating, as a function of $\sigma_*$.}
\label{fig:hgamma}
\end{figure}

In the left panel of Figure \ref{fig:hgamma}, we plot the Hubble parameter and the decay rate as functions of scale factor for several values of $\sigma_*$ near $\sigma_0$. For the smallest $\sigma_*$ values (e.g. the solid blue curve in Figure \ref{fig:hgamma}, left), the decay rate is large enough that the inflaton decays before there is time for the modulus to evolve. As we increase $\sigma_*$ closer to $\sigma_0$ (e.g. dotted red curve in Figure \ref{fig:hgamma}, left), the initial value of $\Gamma(\sigma)$ is smaller and thus the modulus has more time to roll. The modulus rolls away from $\sigma_0$, so the decay rate monotonically increases over time until the inflaton decay takes place at $\Gamma(\sigma)/H=1$. The modulus picks up larger and larger speed over time, so $\Gamma(\sigma)$ increases more rapidly. The decay process is clearly different if we assume that the decay rate is fixed at $\Gamma(\sigma_*)$. In fact, the inflaton would never decay if the modulus is fixed at $\sigma_*=\sigma_0$, which is not the case for a dynamic modulus (dotted red curve in Figure \ref{fig:hgamma}, left). 

For values of $\sigma_*$ slightly larger than $\sigma_0$ (e.g. dashed magenta curve in Figure \ref{fig:hgamma}, left), the decay rate decreases first until the modulus rolls past $\sigma_0$. After $\sigma$ reaches $\sigma_0$, the decay rate increases and soon exceeds $H$ to complete the decay process. However, once $\sigma_*$ is much larger than $\sigma_0$ (e.g long-dashed green curve in Figure \ref{fig:hgamma}, left), the decay condition $\Gamma(\sigma)/H=1$ is satisfied before the modulus reaches $\sigma_0$, and there is suddenly much less time for the modulus to roll. Assuming sudden decay of the inflaton, we therefore find a discontinuity in the function $\mathcal{N}(\sigma_*)$. 

We plot $X(\sigma_{\rm{reh}})$ as a function of $\sigma_*$ on the right
side of Figure \ref{fig:hgamma}. As we increase $\sigma_*$ toward and past $\sigma_0$, the value of $|X(\sigma_{\rm{reh}})|$ also increases until the discontinuity around $\sigma_*=1.033\sigma_0$. The left side of Figure \ref{fig:hgamma} shows that $\dot{\Gamma}$ at reheating is increasing with $\sigma_*$, consistent with the plot of $X(\sigma_{\rm{reh}})\propto-\dot{\Gamma}$ on the right side of Figure \ref{fig:hgamma}. Values of $X(\sigma_{\rm{reh}})\sim1$ are also realized but only within a small range around $\sigma_*=1.033\sigma_0$; for values of $\sigma_*\gtrsim1.04\sigma_0$, the initial decay rate is large enough that the decay rate is nearly constant before reheating (after which the behaviour of $\Gamma$ is irrelevant).

\begin{figure}[t]
\centering
\includegraphics[trim=1.5cm 1.3cm 0.5cm 6.5cm,  width=3.in]{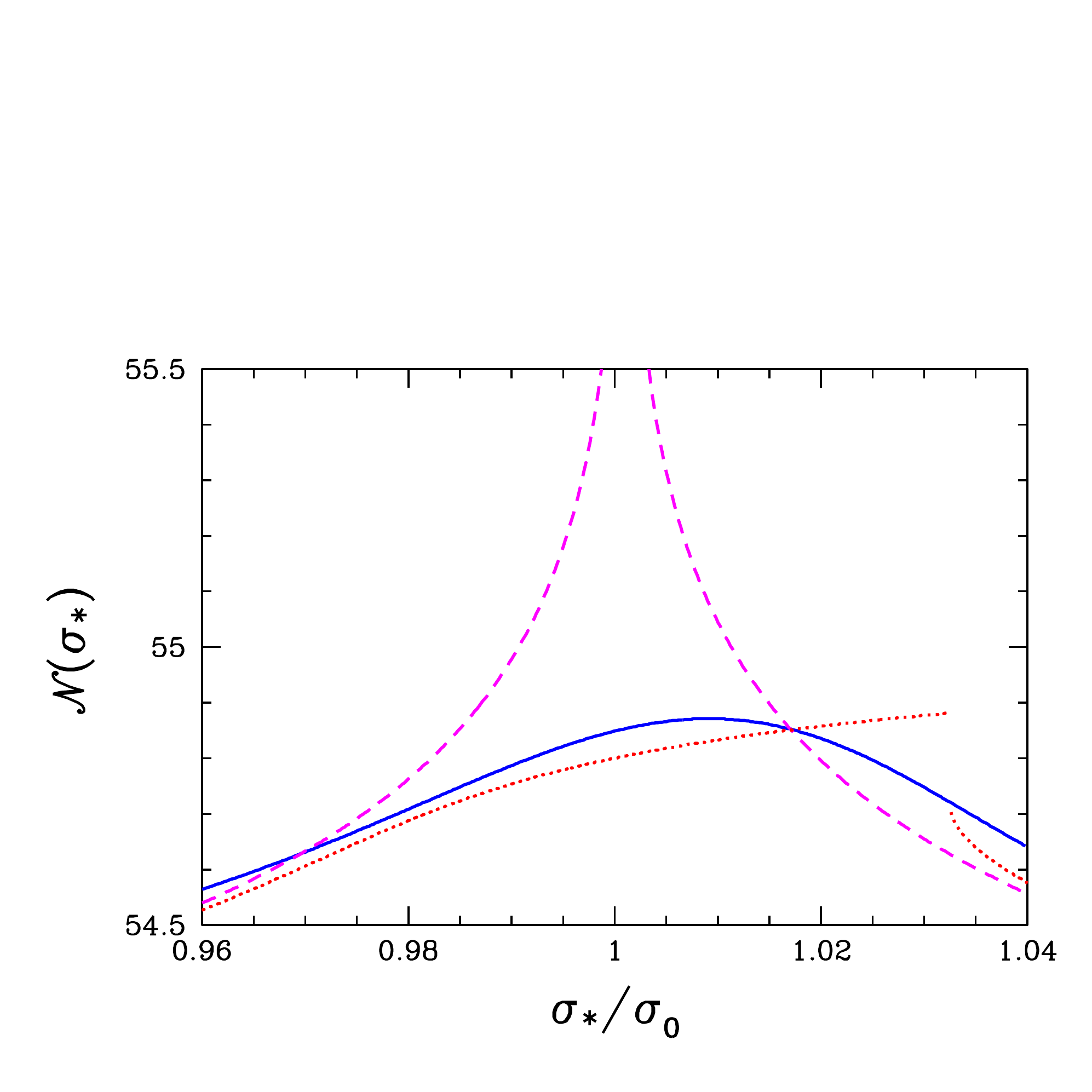}
\includegraphics[trim=0.8cm 1.3cm 1.1cm 209.8cm,  width=3.in]{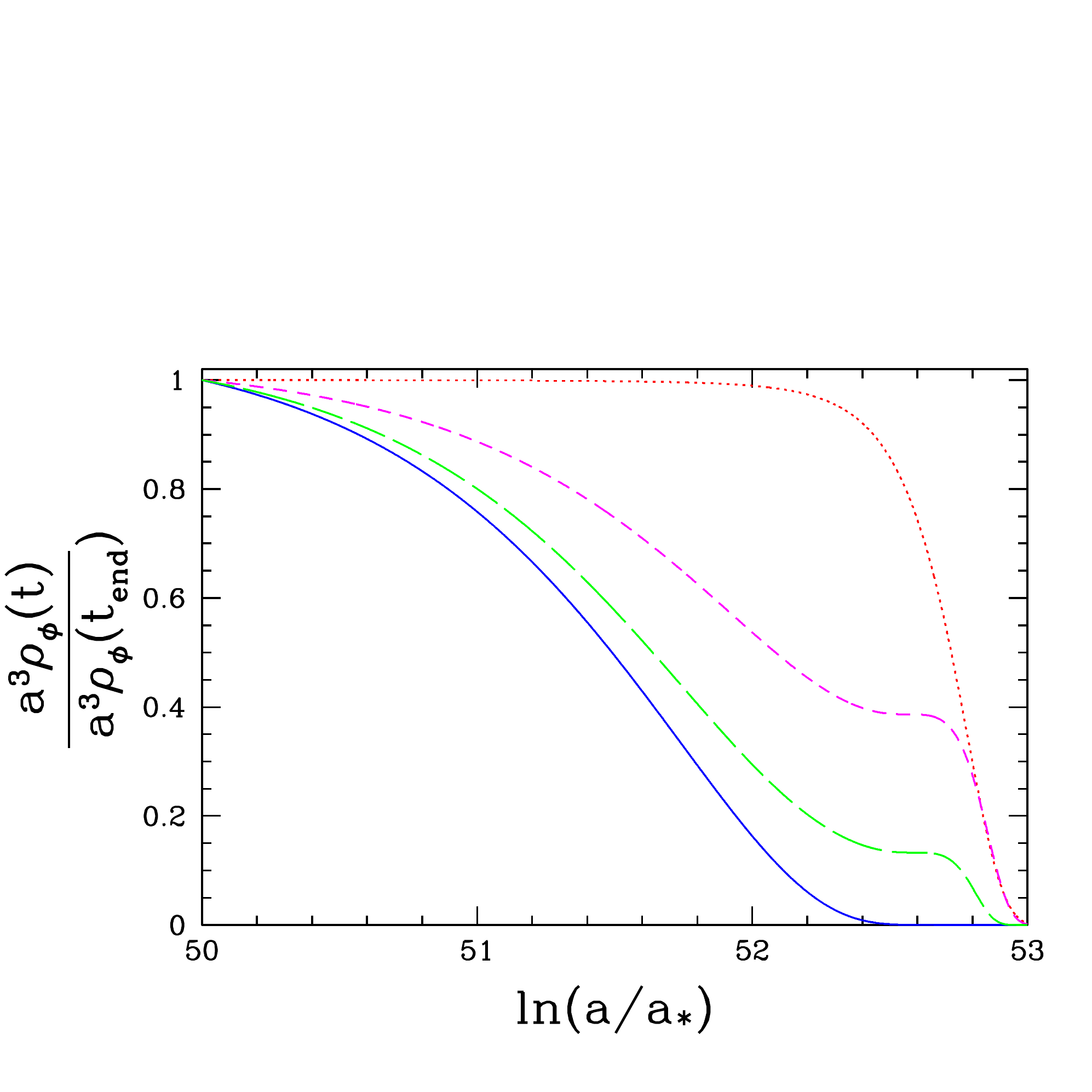}
\caption{Left: the total number of e-foldings between horizon exit and reheating, as a function of $\sigma_*$. The solid blue curve comes from the numerical solution. The dotted red curve is our analytic solution derived assuming sudden decay of the inflaton, while the dashed magenta curve assumes sudden decay as well as a fixed modulus value. Right: the comoving inflaton energy density as a function of scale factor for four values of $\sigma_*$. The colours and line types correspond to those in the left panel of Figure \ref{fig:hgamma}. The energy density is normalized to the inflationary energy density.}
\label{fig:tdec}
\end{figure}

We now compare our analytic solution to the numerical solution, which does not assume sudden decay of the inflaton. We plot $\mathcal{N}(\sigma_*)$ on the left side of Figure \ref{fig:tdec}. The numerical solution is shown by the solid blue curve. Our solution assuming sudden decay is shown in dotted red curve, obtained from Eq.(\ref{efold2}) with $\sigma_{\rm{reh}}$ given by Eq.(\ref{s*sdec}). We have chosen $\rho_f=10^{52}\,\rm{GeV}^4$ for the final hypersurface, selected to satisfy $\rho_{\phi}(t_f)\ll\rho_\gamma(t_f)$. Any other value for $\rho_f$ in the radiation-dominated Universe will only shift $\mathcal{N}$ by a constant, which does not affect the observables. We also overlay a dashed magenta curve showing $\mathcal{N}(\sigma_*)$ if the modulus field is fixed at $\sigma_*$, which is obtained from Eq.(\ref{efold2}) with $\sigma_{\rm{reh}}=\sigma_*$.

Our analytic solution with rolling effects correctly predicts that the modulus's evolution makes $\mathcal{N}(\sigma_*)$ less sensitive to $\sigma_*$ when $\sigma_*\simeq\sigma_0$. As we have discussed, we find that this solution has a discontinuity near $\sigma_*=1.033\sigma_0$, while the numerical solution is smooth and takes a maximum point near $\sigma_*=1.007\sigma_0$. The failure of the sudden-decay approximation, and consequently of our analytic solution, arises because the decay process is changing in a non-trivial way, which is clear from the shape of $\Gamma(\sigma(t))$ in Figure \ref{fig:hgamma} (left panel). 
Since the decay rate initially decreases for values of $\sigma_*>\sigma_0$, there can be two stages of inflaton decay. On the right side of Figure \ref{fig:tdec}, we show the time evolution of the comoving inflaton energy density, $a^3(t)\rho_{\phi}(t)$, normalized to its value at the end of inflation.  We clearly see that there are two decay phases for $\sigma_*>\sigma_0$, unless $\sigma_*$ is large enough that the inflaton energy density nearly vanishes during the first decay phase.

\begin{figure}[t]
\centering
\includegraphics[trim=1.4cm 1.3cm 0.5cm 6.5cm,  width=3.in]{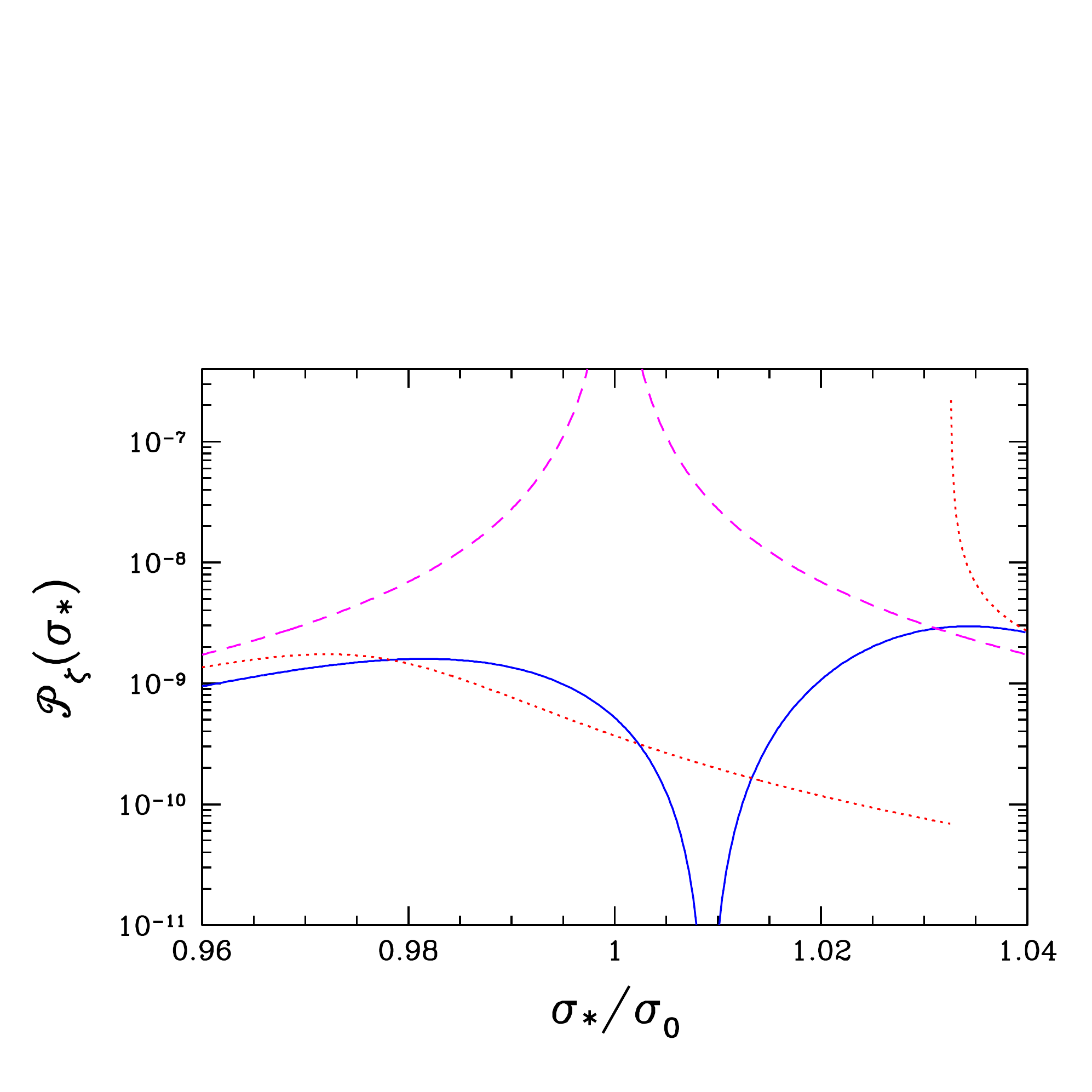}
\includegraphics[trim=1.4cm 1.3cm 0.5cm 6.5cm,  width=3.in]{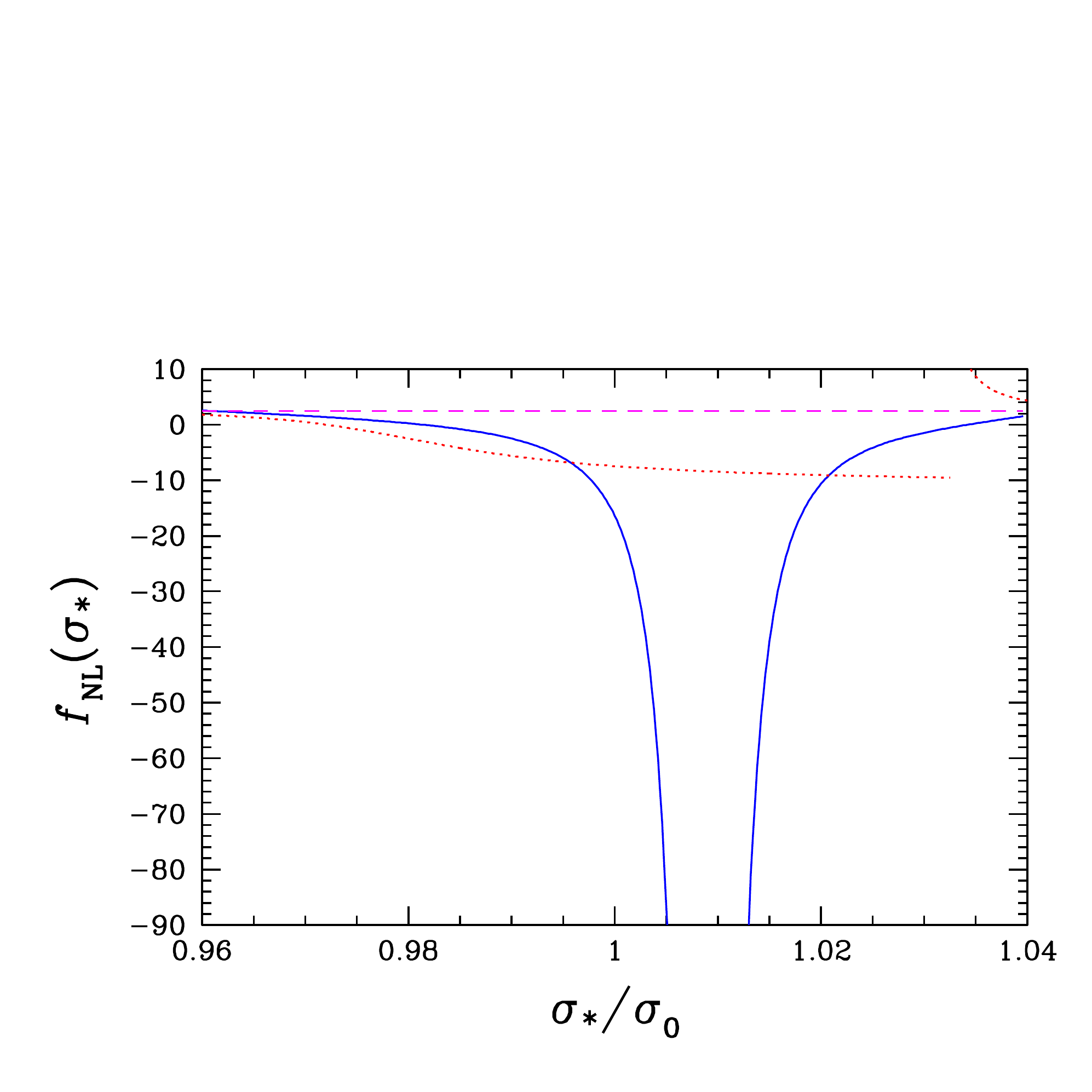}
\caption{The power spectrum (left) and $f_\mathrm{NL}$ (right) as functions of $\sigma_*$. The solid blue curve comes from the numerical solution. The dotted red curve is our analytic solution derived assuming sudden decay of the inflaton, while the dashed magenta curve assumes sudden decay as well as a fixed modulus value.}
\label{fig:PfNL}
\end{figure}

It is clear from the left panel of Figure \ref{fig:tdec} that the slope of $\mathcal{N}(\sigma_*)$ for the analytic solution with rolling effects, analytic solution without rolling effects, and numerical solution all differ, and consequently the observables behave differently as well. We plot $\mathcal{P}_{\zeta}(\sigma_*)$ and $f_{\rm{NL}}(\sigma_*)$ in Figure \ref{fig:PfNL}. Our solution assuming the sudden inflaton decay at $\Gamma(\sigma)/H=1$ is shown in dotted red curve, while the numerical solution without assuming sudden decay is the solid blue curve. We also overlay a dashed, magenta curve assuming that the modulus is frozen at $\sigma_*$; the power spectrum is Eq.(\ref{Pinvno}) with $n=2$, and $f_{\rm{NL}}=2.5$. Our analytic solution including rolling effects predicts that the e-folding number near $\sigma_*=\sigma_0$ becomes less sensitive to $\sigma_*$ due to the evolution of the modulus, and consequently $\mathcal{P}_{\zeta}(\sigma_*)$ is suppressed. The power spectrum according to the numerical solution is similarly suppressed but it vanishes completely at $\sigma_*=1.007\sigma_0$, as $\mathcal{N}(\sigma_*)$ takes a maximum at this point. The analytic solution with rolling effects fails to accurately predict $\mathcal{P}_{\zeta}(\sigma_*)$ because it approximates the complicated decay process (seen in the right panel of Figure \ref{fig:tdec}) by an instant decay. Even though the lightness condition at reheating $m^2\ll\Gamma(\sigma_{\rm{reh}})^2$ is also only marginally satisfied for $\sigma_*\simeq\sigma_0$, the discrepancy we see in Figure \ref{fig:PfNL} between the analytical prediction with rolling effects and numerical solution comes almost exclusively from the failure of the sudden-decay approximation. The analytic solution with rolling does however correctly predict that the power spectrum can be overestimated by many orders of magnitude if the modulus rolling is neglected. 

The right panel of Figure \ref{fig:PfNL} shows that $f_\mathrm{NL}$ according to the analytic solution with rolling asymptotes to $f_{\rm{NL}}(\sigma_*)\simeq-10$ as the value of $|X(\sigma_{\rm{reh}})|$ in Figure \ref{fig:hgamma} (right panel) increases with $\sigma_*$ until the discontinuity. A positive correction to $f_{\rm{NL}}(\sigma_*)$ is observed right after the discontinuity, increasing it from 2.5 to $f_{\rm{NL}}(\sigma_*)\gtrsim10$. However, comparing this analytic solution to the numerical solution, we see that this result is an artifact introduced by the sudden-decay approximation. We further see that the non-linearity parameter according to the numerical solution takes much larger negative values near the maximum point of $\mathcal{N}(\sigma_*)$, and is undefined at the maximum of $\mathcal{N}(\sigma_*)$ (which is consistent with the fact that the power spectrum vanishes here). 

This example illustrates the conditions under which one should expect the sudden-decay approximation to be inaccurate. The sudden-decay approximation characterizes the entire decay process as an instantaneous transition that occurs when $\Gamma\simeq H$. It therefore fails to capture any deviation of $\rho_{\phi}(t)$ from $\rho_\phi \propto a^{-3}$ before the sudden transition and $\rho_\phi= 0$ after the transition.  If $\Gamma(\sigma(t))/H(t)$ decreases before or shortly after $\Gamma\simeq H$, then the inflaton decays in multiple stages (see Figure \ref{fig:tdec}), and reheating cannot be accurately described by a single transition.  In the model considered in this section, the decay rate with values of $\sigma_*$ in the range $\sigma_0<\sigma_*\lesssim1.04\sigma_0$ is non-monotonic prior to reheating, and thus we observe discrepancies between the formulas we derived in section 2.1 and the numerical solution, which does not assume sudden decay.

\section{Models with a Varying Effective Modulus Mass}
\FloatBarrier

We have shown in the previous sections that, for a quadratic potential, the importance of modulus dynamics can be measured by one quantity, $X(\sigma_{\rm{reh}})$; the modulus dynamics become relevant when $|X(\sigma_{\rm{reh}})|\gtrsim1$. However, we have also seen in section 2.2 that including the rolling effects of the modulus introduces an additional term to $f_{\rm{NL}}$ that is proportional to $X(\sigma_{\rm{reh}})^{-1}$ [see Eq.(\ref{Xsmall})]. This term vanishes if the effective mass of the modulus is constant. In order to study the significance of this term, we now consider a quartic potential, 
\begin{equation} \label{Vquart}
V(\sigma)=\frac{1}{4}\eta\sigma^4,
\end{equation}
with model parameters that satisfy $|X(\sigma_{\rm{reh}})|\ll1$. 

\begin{figure}[t]
\centering
\includegraphics[width=3.1in]{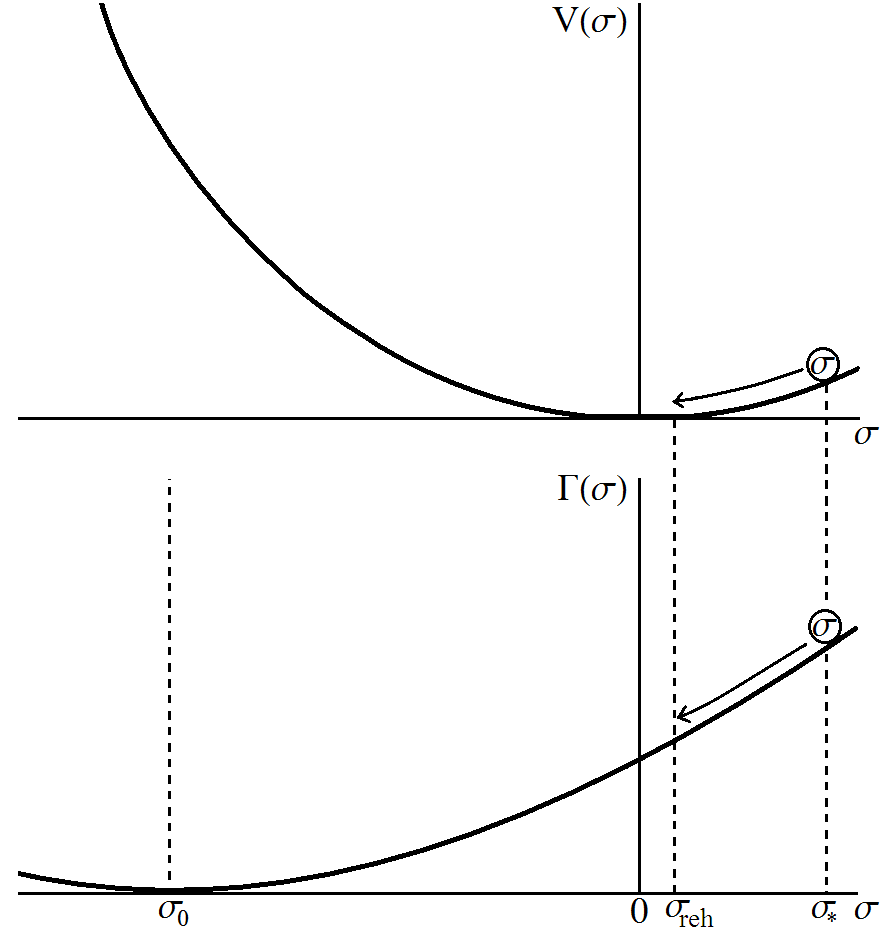}
\caption{Illustration of the quartic modulus potential $V(\sigma)$ and the inflaton decay rate $\Gamma(\sigma)$ for the model in section 4. The inflaton decay rate is a quadratic function of $\sigma$ with a shift $\sigma_0$ of the minimum point. The modulus initially has some positive field value satisfying $|\sigma_0|\gg\sigma_*>0$, and it rolls towards $\sigma=0$.}
\label{fig:quart}
\end{figure}

We first show that the power spectrum is not significantly affected by the evolution of the modulus in the limit of small $|X(\sigma_{\rm{reh}})|$. The general formula Eq.(\ref{P3}) for $\mathcal{P}_{\zeta}$, compared to Eq.(\ref{P3no}) which neglects the modulus dynamics, contains a factor of 
\begin{equation} \label{Pcorr}
\left[\frac{1}{1-X(\sigma_{\rm{reh}})}\frac{V'(\sigma_{\rm{reh}})}{V'(\sigma_*)}\right]^2=\left(\frac{\sigma_{\rm{reh}}}{\sigma_*}\right)^6\left[1+\mathcal{O}(X(\sigma_{\rm{reh}}))\right].
\end{equation}
Solving the approximate EOM for the modulus Eq.(\ref{cEOM}) with the
quartic potential Eq.(\ref{Vquart}) (assuming a constant~$H$ during
inflation), we obtain 
\begin{equation} \label{sigmaquart}
\sigma(t)^2=\sigma_*^2\left[1+\frac{4}{27}\frac{\eta\sigma_*^2}{H_*^2}\left(\frac{9}{2}\mathcal{N}_*-1+\frac{H_*^2}{H(t)^2}\right)\right]^{-1}
\end{equation}
which, upon evaluation at $t=t_{\rm{reh}}$ defined by $\Gamma(\sigma(t_{\rm{reh}}))/H(t_{\rm{reh}})=\beta$, can be used to show that  
\begin{equation} \label{s*sdecquart2}
\left(\frac{\sigma_{\rm{reh}}}{\sigma_*}\right)^2=\left[1+\frac{4}{27}\frac{\eta\sigma_*^2}{H_*^2}\left(\frac{9}{2}\mathcal{N}_*-1\right)\right]^{-1}\left[1-\frac{4\beta^2}{27}\frac{\eta\sigma_{\rm{reh}}^2}{\Gamma(\sigma_{\rm{reh}})^2}\right]=1+\mathcal{O}\left(\frac{V''(\sigma_*)}{H_*^2},\frac{V''(\sigma_{\rm{reh}})}{\Gamma(\sigma_{\rm{reh}})^2}\right). 
\end{equation}
By using Eq.(\ref{cEOM}), we have already assumed that $V''(\sigma_*)/H_*^2$ and $V''(\sigma_{\rm{reh}})/\Gamma(\sigma_{\rm{reh}})^2$ are small compared to unity, and now we assume that $\mathcal{N}_*\times V''(\sigma_*)/H_*^2$ is small as well.
We conclude from Eqs.(\ref{Pcorr}, \ref{s*sdecquart2}) that the correction to the power spectrum introduced by modulus dynamics is limited by the size of $V''/H^2\ll1$ and $|X(\sigma_{\rm{reh}})|\ll1$. This conclusion can be generalized to all power-law potentials $V\propto\sigma^p$.

The non-linearity parameter assuming a quartic potential and $|X(\sigma_{\rm{reh}})|\ll1$ is [see Eq.(\ref{Xsmall})]
\begin{align} \label{fNLquart}
f_{\rm{NL}}=&5\left\{1-\frac{\Gamma(\sigma_{\rm{reh}})\Gamma''(\sigma_{\rm{reh}})}{\Gamma'(\sigma_{\rm{reh}})^2}\left[1+\mathcal{O}(X(\sigma_{\rm{reh}}))\right]-3\frac{\Gamma(\sigma_{\rm{reh}})}{\Gamma'(\sigma_{\rm{reh}})\sigma_{\rm{reh}}}\left(1-\frac{\sigma_*^2}{\sigma_{\rm{reh}}^2}\right)\right\} \nonumber\\
&+\mathcal{O}\left(X(\sigma_{\rm{reh}}),\frac{V''(\sigma_{\rm{reh}})}{\Gamma(\sigma_{\rm{reh}})^2},\frac{V''(\sigma_*)}{\Gamma(\sigma_{\rm{reh}})^2}\right).
\end{align}
Among the terms arising from the modulus's evolution, the last term in
the first line can dominate over 
$5 (1 - \Gamma \Gamma'' / \Gamma'^2)$ and give contributions to
$f_{\mathrm{NL}}$ of order unity or more.
We denote this term by $\Delta f_{\rm{NL}}$: 
\begin{equation} \label{DfNL}
\Delta f_{\rm{NL}}=-15\frac{\Gamma(\sigma_{\rm{reh}})}{\Gamma'(\sigma_{\rm{reh}})\sigma_{\rm{reh}}}\left(1-\frac{\sigma_*^2}{\sigma_{\rm{reh}}^2}\right).
\end{equation}
This component of $f_{\rm{NL}}$ will be important when $\Gamma(\sigma)$ does not depend sensitively on $\sigma$ so that $|\Gamma'(\sigma_{\rm{reh}})\sigma_{\rm{reh}}/\Gamma(\sigma_{\rm{reh}})|\ll1$. Note that if the decay rate is a power law $\Gamma(\sigma)\propto\sigma^n$ that shares its minimum point with $V(\sigma)$, then $\Delta f_{\rm{NL}}$ will be of order $V''/H^2\ll1$ as seen in Eq.(\ref{s*sdecquart2}) (except for the special case $|n|\ll1$). In the following, we assume a decay rate of the form\footnotemark\footnotetext{Another interesting example one can consider is a decay rate of the form $\Gamma(\sigma)\propto e^{-\sigma/M_{\rm{PL}}}$ with $\sigma\ll M_{\rm{PL}}$, which arises from e.g. dilatonic couplings. Without the modulus dynamics one would have $f_{\rm{NL}}=0$ in such a class of models [see Eq.(\ref{fNL3no})], but $|\Delta f_{\rm{NL}}|$ can be large since $\left|\frac{\Gamma'(\sigma_{\rm{reh}})\sigma_{\rm{reh}}}{\Gamma(\sigma_{\rm{reh}})}\right|=\left|\frac{\sigma_{\rm{reh}}}{M_{\rm{PL}}}\right|\ll1$.} 
\begin{equation} \label{gamma3}
\Gamma(\sigma)=\mu^{-1}(\sigma-\sigma_0)^2.
\end{equation}
We study a parameter space that satisfies
$|\sigma_{\rm{reh}}-\sigma_0|\gg|\sigma_{\rm{reh}}|$ so that
$|\Gamma'(\sigma_{\rm{reh}})\sigma_{\rm{reh}}/\Gamma(\sigma_{\rm{reh}})|\ll1$. By writing $X(\sigma_{\rm{reh}})$ as in Eq.(\ref{Xpower}), we see that such a nearly constant decay rate leads to $|X(\sigma_{\rm{reh}})|\ll1$ for power law potentials. In this model, $\Gamma(\sigma)$ does not vary significantly prior to reheating: $\Gamma(\sigma_{\rm{reh}}) = \Gamma(\sigma_*) \times(1+\mathcal{O}(V''/H^2))$.  Therefore, we may evaluate Eq.(\ref{sigmaquart}) at $t=t_{\rm{reh}}$ and
replace $\Gamma(\sigma_{\rm{reh}})$ by $\Gamma(\sigma_*)$ to obtain 
the following approximation up to $\mathcal{O}(V''/H^2)$: 
\begin{equation} \label{mess2}
\sigma_{\rm{reh}}=\sigma_*\left[1-\frac{2}{27}\frac{\eta\sigma_*^2}{H_*^2}\left(\frac{9}{2}\mathcal{N}_*-1\right)-\frac{2\beta^2}{27}\frac{\eta\mu^2\sigma_*^2}{(\sigma_*-\sigma_0)^4}\right],
\end{equation}
which allows us to write $\mathcal{P}_{\zeta}(\sigma_*)$ and $f_{\rm{NL}}(\sigma_*)$ analytically in terms of the model parameters. 

\begin{figure}[t]
\centering
\includegraphics[trim=1.3cm 1.3cm 0.5cm 6.5cm,  width=3.in]{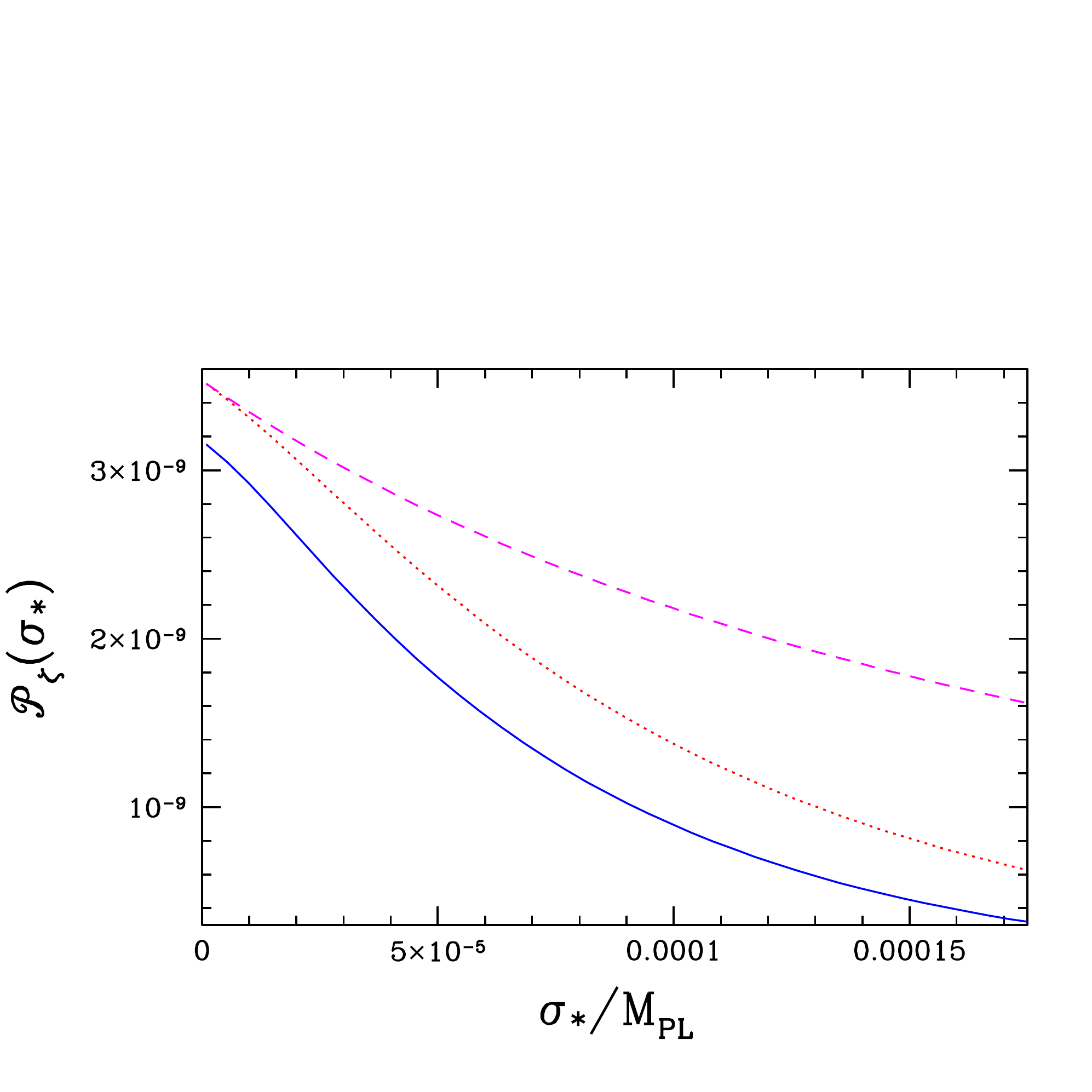}
\includegraphics[trim=1.5cm 1.3cm 0.5cm 6.5cm,  width=3.in]{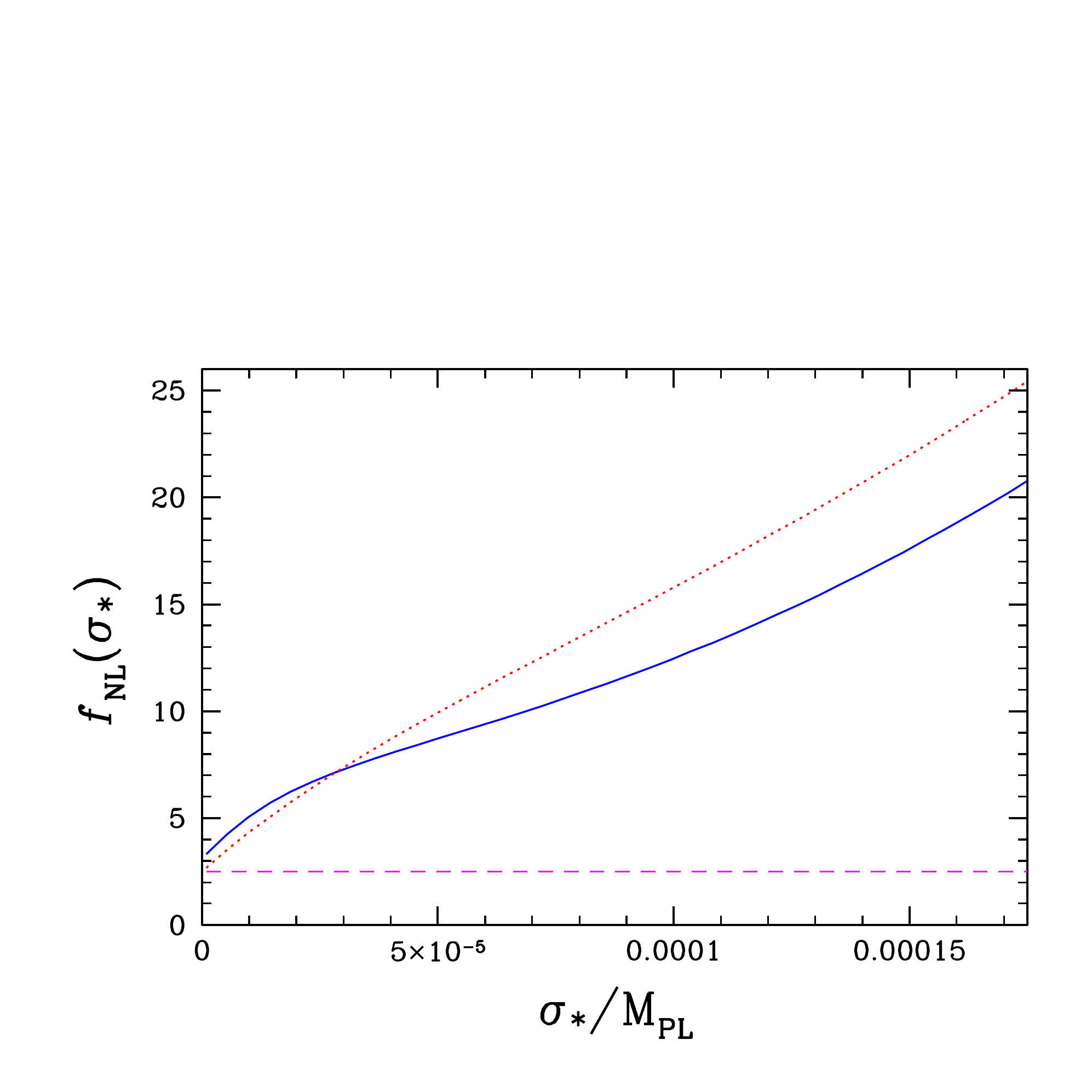}
\caption{The power spectrum (left) and the non-linearity parameter (right) as functions of $\sigma_*$. The solid blue curve comes from the numerical solution. The dotted red curve is our analytic solution derived assuming sudden decay of the inflaton, while the dashed magenta curve assumes sudden decay as well as a fixed modulus value.}
\label{fig:PfNLquart}
\end{figure}

We now study a specific model in which $|\Delta f_{\rm{NL}}|$ takes large values and thus $|f_{\rm{NL}}|\gg1$. We take the following set of parameters as a definite example: \{$H_*=10^{12}\rm{GeV}$, $\mathcal{N}_*=50$, $\eta=5.8\times10^{-8}$, $\mu=7.3M_{\rm{PL}}$, $\sigma_{0}=-3.7\times10^{-4}M_{\rm{PL}}\}$. 
We again set $\beta=1$ in the following calculations; the effects of changing $\beta$ are not large compared to the discrepancy between the analytic and numerical results (shown in Figure \ref{fig:PfNLquart}), and thus the specific value of $\beta$ has little effect on the following analysis. We have chosen $\sigma_0<0$ and we consider $\sigma_*$ values satisfying $|\sigma_0|\gg\sigma_*>0$ such that the decay rate slowly decreases in time. An illustration of the setup is provided in Figure \ref{fig:quart}. We note that we can similarly construct a slowly increasing $\Gamma(\sigma(t))$ by switching the sign of $\sigma_0$, which also changes the sign of $\Delta f_{\rm{NL}}$ as seen in Eq.(\ref{DfNL}).

\begin{figure}[t]
\centering
\includegraphics[trim=1.2cm 1.3cm 0.5cm 6.5cm,  width=3.in]{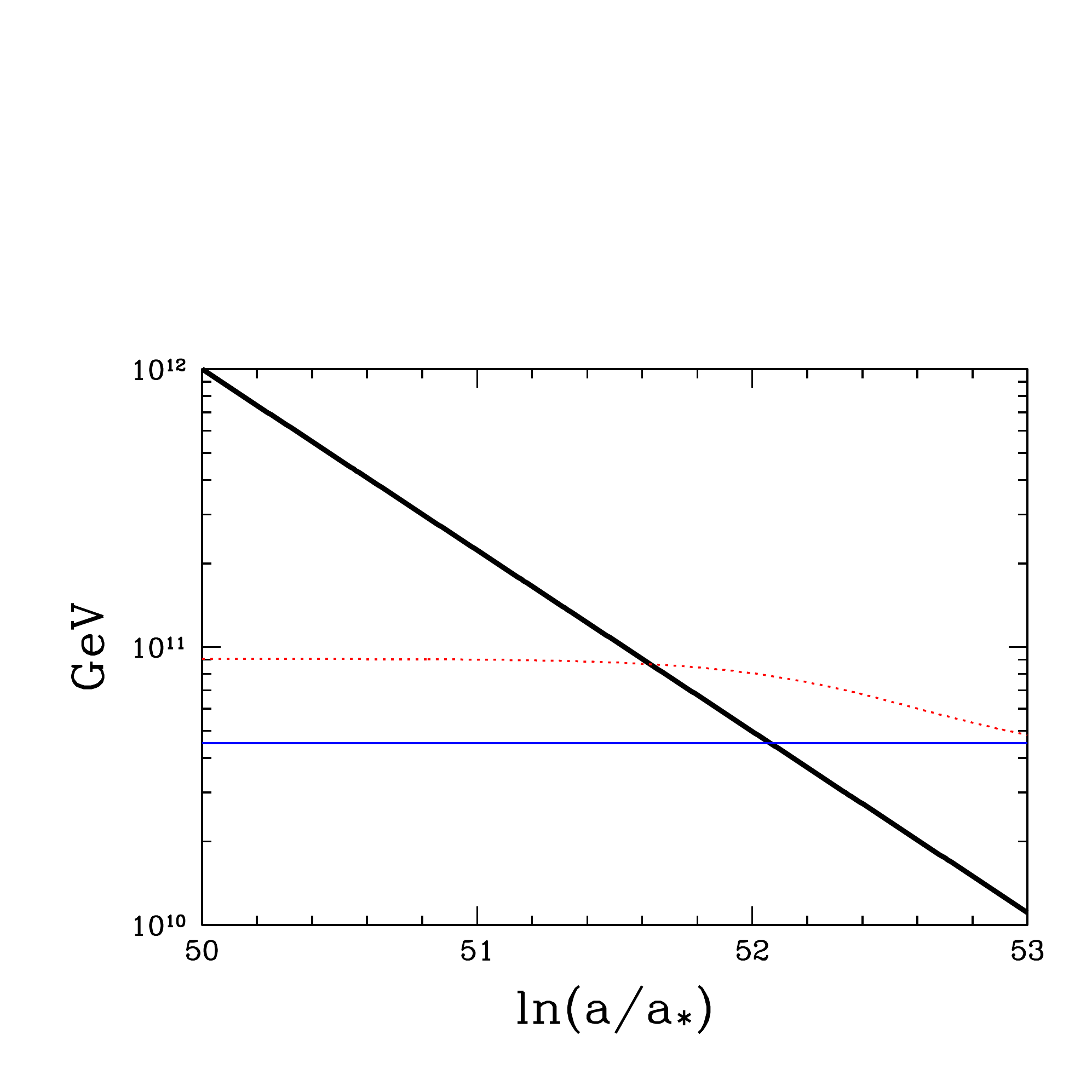}
\includegraphics[trim=0.8cm 1.3cm 0.9cm 6.5cm,  width=3.in]{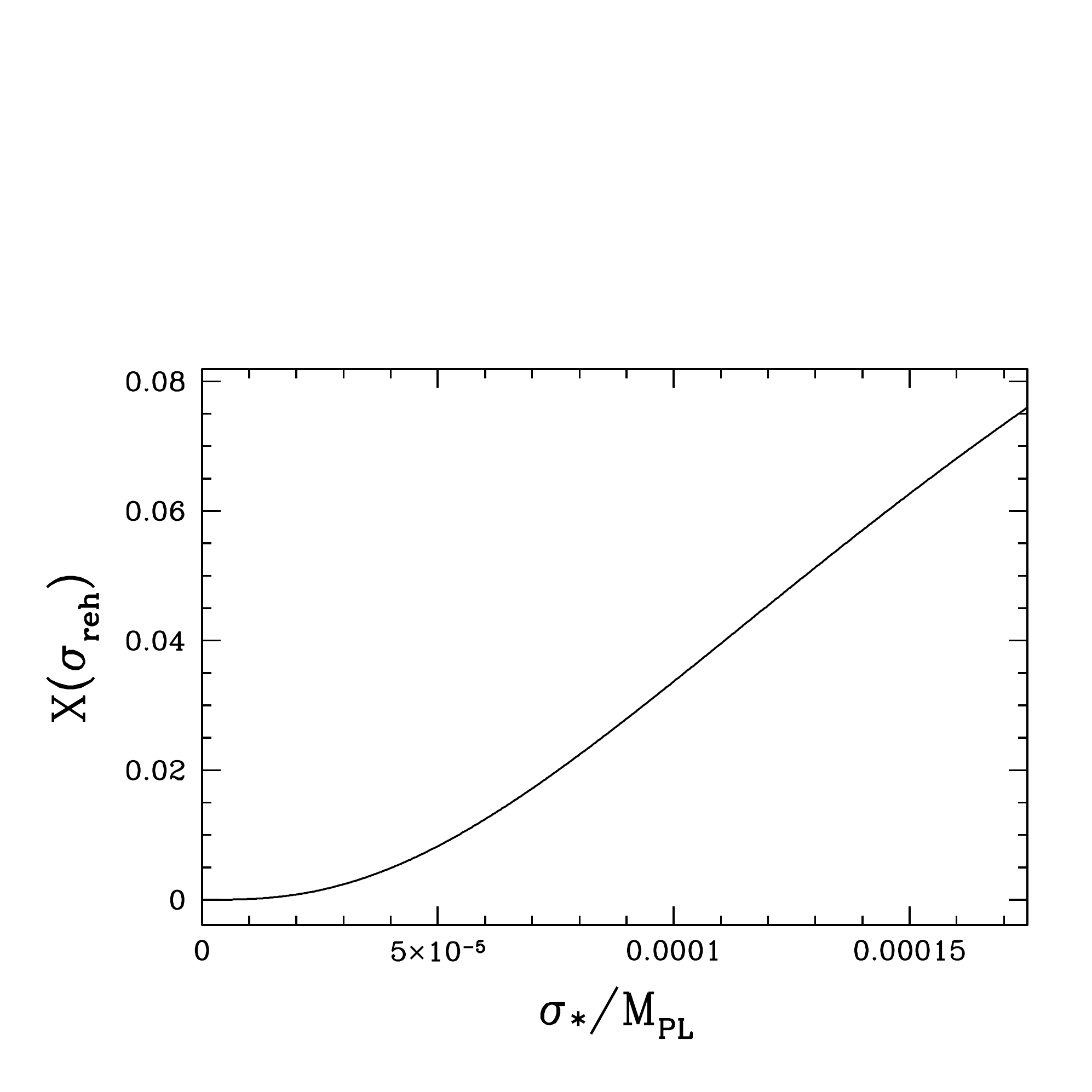}
\caption{Left: plot of the Hubble parameter (thick black line) and the decay rate as a function of scale factor for two values of $\sigma_*$: $0$ (in solid blue) and $1.8*10^{-4}M_{\rm{PL}}$ (in dotted red). These two values encompass the range of $\sigma_*$ values that we consider. Right: Plot of $X(\sigma_{\rm{reh}})$, a measure of $\dot\Gamma$ at reheating, as a function of~$\sigma_*$.}
\label{fig:hgammaquart}
\end{figure}

The effective modulus mass $V''(\sigma)=3\eta\sigma^2$ does not stay
constant for a quartic potential. The modulus thus stays light for a
different amount of time depending on $\sigma_*$; larger $\sigma_*$
implies larger effective mass and so the lightness condition is violated
closer to the time when inflation ends. For the given parameters, the
range of $\sigma_*$ for which the modulus stays light until reheating is
$\sigma_*<1.8\times  10^{-4}M_{\rm{PL}}$. 
In the left panel of Figure \ref{fig:hgammaquart}, we plot the Hubble rate and the decay rate as functions of scale factor. We plot the decay rate with initial values $\sigma_*=0$ and $\sigma_*=1.8\times10^{-4}M_{\rm{PL}}$, which are the boundary points of the range of $\sigma_*$ that we consider. 
The decay rates for the two initial values are nearly constant prior to reheating. As we have already discussed, a nearly constant $\Gamma$ during reheating implies $|X(\sigma_{\rm{reh}})|\ll1$, which we confirm in the right panel of Figure \ref{fig:hgammaquart}.

We present the power spectrum and the non-linearity parameter in Figure \ref{fig:PfNLquart}. The solid blue line corresponds to the numerical solution, which does not assume sudden decay of the inflaton. The dotted red curve is our analytic solution, which we obtain from Eqs.(\ref{P3}, \ref{fNL3}) with $\sigma_{\rm{reh}}$ given by Eq.(\ref{mess2}). The dashed magenta curve is the analytic solution neglecting the modulus dynamics, given by Eq.(\ref{Pinvno}) for $\mathcal{P}_{\zeta}(\sigma_*)$ with $n=2$, and $f_{\rm{NL}}=2.5$. 
The numerical solution reveals that the rolling of the modulus suppresses the amplitude of power spectrum and makes $f_{\rm{NL}}(\sigma_*)$ monotonically increase with $\sigma_*$ instead of being fixed at $2.5$.  For this parameter set, non-Gaussianity is enhanced up to $f_{\rm{NL}}\simeq20$, which far exceeds the upper bound established by current observations. The analytic solution including the rolling effects correctly predicts these features, but Figure \ref{fig:PfNLquart} shows that it does not exactly match the numerical solution.  Furthermore, as we discussed at the beginning of the section, analytic calculations in the limit of small $|X(\sigma_{\rm{reh}})|$ predict that $\mathcal{P}_{\zeta}(\sigma_*)$ should not be significantly affected by the modulus's evolution, but both the numerical and analytical solutions shown in Figure \ref{fig:PfNLquart} indicate that the rolling of the modulus suppresses $\mathcal{P}_{\zeta}(\sigma_*)$ by up to a factor of $\sim\!5$.  Both of these discrepancies result from a breakdown of the approximations we used in our derivations. While deriving our analytic results including the rolling effects, we assumed that the sudden-decay approximation is applicable, which requires $\Gamma(\sigma_{\rm{reh}})\ll H_*$ (as we pointed out in section 3.1), that the modulus is light, and also that $\mathcal{N}_*\times V''(\sigma_*)/H_*^2\ll1$ [which is assumed following Eq.(\ref{s*sdecquart2})]. For $\sigma_*\simeq1.8\times  10^{-4}M_{\rm{PL}}$ in Figure \ref{fig:PfNLquart}, however, we employed $\Gamma(\sigma_{\rm{reh}})/H_*\sim10^{-1}$, $V''(\sigma_{\rm{reh}})/\Gamma(\sigma_{\rm{reh}})^2\sim1$, and $\mathcal{N}_*\times V''(\sigma_*)/H_*^2\sim1$.  

We were forced to choose these parameters because, for a quartic potential, parameter sets with $|X(\sigma_{\rm{reh}})|\ll1$ that result in $|\Delta f_{\rm{NL}}|\gg1$ at least marginally violate the assumptions required for the analytic calculations. To prove this point, we first use the power spectrum formula Eq.(\ref{P3}) to write $\Delta f_{\rm{NL}}$ in Eq.(\ref{DfNL}) as 
\begin{equation} 
|\Delta f_{\rm{NL}}|=\left|\frac{5}{4\pi}\mathcal{P}_{\zeta}^{-1/2}\frac{1}{1-X(\sigma_{\rm{reh}})}\frac{H_*\sigma_{\rm{reh}}^2}{\sigma_*^3}\left(1-\frac{\sigma_*^2}{\sigma_{\rm{reh}}^2}\right) \right|.
\end{equation} 
Further requiring that during inflation the modulus's classical rolling
dominates over the quantum fluctuations [as in Eq.(\ref{stoc})], 
\begin{equation} \label{stoc2}
 \frac{H_*}{2\pi} \ll \frac{|\dot{\sigma}_*|}{H_*} = \left|\frac{\eta\sigma_*^3}{3H_*^2} \right|,
\end{equation}
and assuming that the modulus is light at reheating, $3\eta\sigma_{\rm{reh}}^2\ll\Gamma(\sigma_{\rm{reh}})^2$, we obtain
\begin{equation}
|\Delta f_{\rm{NL}}|\ll \left|\frac{5}{18}\mathcal{P}_{\zeta}^{-1/2}\frac{1}{1-X(\sigma_{\rm{reh}})}\frac{\Gamma(\sigma_{\rm{reh}})^2}{H_*^2}\left(1-\frac{\sigma_*^2}{\sigma_{\rm{reh}}^2}\right)\right|.
\end{equation} 
The last term in the brackets is smaller than unity for a light modulus [see Eq.(\ref{s*sdecquart2})], and the sudden-decay approximation is applicable only if
$\Gamma(\sigma_{\rm{reh}})\ll H_*$. Setting $\mathcal{P}_{\zeta}$ to the observed amplitude of
$2.2\times10^{-9}$ and $\Gamma(\sigma_{\rm{reh}})/H_*=10^{-2}$ as an
example, then for $|X(\sigma_{\mathrm{reh}})| \ll 1$ we obtain $|\Delta
f_{\rm{NL}}|\ll1$.  In order to significantly affect $f_{\rm{NL}}$ with modulus dynamics while keeping $\mathcal{P}_{\zeta}$ near the observed value, a model must at least marginally violate the assumptions required for the analytic calculations, like the model used to generate Figure \ref{fig:PfNLquart}.\footnotemark\footnotetext{In order to illustrate the behavior of the system, we have included $\sigma_* \simeq 0$ in the figures. We should remark that Eq.(\ref{stoc2}) does not hold for $\sigma_* / M_{\rm{PL}} \lesssim 8\times10^{-5}$, so it is questionable to assume that $\sigma$ evolves classically for these values of $\sigma_*$.  However, we also note that deviations from classical evolution during inflation may have little effect on the resulting perturbations, since it is the modulus's evolution after inflation that significantly affects the inhomogeneous reheating process.}
This conclusion can be generalized to any power-law modulus potential $V\propto\sigma^p$:
$|f_{\rm{NL}}|\gg1$ from modulated reheating scenarios with $|X(\sigma_{\mathrm{reh}})| \ll1$ implies that either the decay rate satisfies $|\Gamma(\sigma_{\rm{reh}})\Gamma''(\sigma_{\rm{reh}})/\Gamma'(\sigma_{\rm{reh}})^2|\gg1$, $\mathcal{P}_\zeta$ is smaller than the observed value, or the conditions assumed for the analytic analyses [such as the lightness condition and Eq.(\ref{stoc2})] are marginally violated.
However, we note that there may be more complicated potentials for which this conclusion does not hold. 

\section{Discussion}
Density perturbations in the modulated reheating scenario are generated from inhomogeneous decay of the inflaton field. The inflaton decay rate $\Gamma$ is controlled by a light field $\sigma$, which acquires fluctuations during inflation and thus modulates the inflaton decay rate. We studied the generation of curvature perturbations in this scenario while monitoring the evolution of the modulus field. Our key results are the expressions for the power spectrum $\mathcal{P}_{\zeta}$ given by Eq.(\ref{P3}) and the non-linearity parameter $f_{\rm{NL}}$ in Eq.(\ref{fNL3}) of curvature perturbations from modulated reheating.  By comparing these expressions to the previously derived results, given by Eqs.(\ref{P3no}, \ref{fNL3no}), which assumed a static modulus, we showed that the dynamics of the modulus field can drastically modify the amplitude of $\mathcal{P}_{\zeta}$ and $f_{\rm{NL}}$. 

The corrections introduced to $\mathcal{P}_{\zeta}$ and $f_{\rm{NL}}$ by the evolution of the modulus can be significant even if the effective modulus mass is small compared to the Hubble rate. Our analytic calculations in section 2 predict that the modulus's evolution will be important under two general conditions: when the inflaton decay rate rapidly changes during reheating, and when the effective mass of the modulus $V''(\sigma)$ varies between inflation and reheating. If $\Gamma(\sigma)$ increases rapidly during reheating, we generically find $f_{\rm{NL}}\simeq-10$ due to the modulus dynamics, and the rolling of the modulus can also significantly suppress the amplitude of the primordial power spectrum.  Even if the decay rate is nearly constant, the modulus dynamics can enhance $|f_{\rm{NL}}|$ by an order of magnitude or more if the effective mass of the modulus is not constant.  These predictions of our analytic formulas were confirmed by numerically analyzing specific models in sections 3 and 4. Statistics of the perturbations calculated from specific models studied previously\cite{zaldarriaga04, bartolo04, vernizzi04, suyama08, ichikawa08, elliston13} may also be greatly altered by fully accounting for the modulus's evolution, but its significance depends on how the modulus potential is specified. Our formalism lets us calculate the curvature perturbation including the effects of the modulus dynamics in any model, given the modulus potential and inflaton decay rate.

Our main expressions Eqs.(\ref{P3}, \ref{fNL3}) for $\mathcal{P}_{\zeta}$ and $f_{\rm{NL}}$ assume that the effective modulus mass is small relative to the Hubble rate, $|V''(\sigma)|\ll H^2$. The modulus field has to be light during inflation to acquire fluctuations, but this inequality does not necessarily need to hold after inflation. We can expect that the modulus dynamics will affect the curvature perturbation more significantly if the Hubble rate reaches values below $|V''(\sigma)|$ before reheating. We confirmed numerically that making $V''(\sigma)\gtrsim H^2$ during reheating can suppress the power spectrum by many orders of magnitude and enhance non-Gaussianity from $f_{\rm{NL}}\simeq-10$ to $f_{\rm{NL}}=\mathcal{O}(-100)$. 

In our analytic calculations, we also treated the inflaton's decay as a sudden event that instantaneously converted all the energy density in the inflaton field to radiation.
The use of this approximation is not warranted if $\Gamma$ decreases during reheating, leading to a decay process that does not proceed rapidly in one step. By studying a model in which the inflaton decays in two stages, we discovered that the evolution of the modulus can modify the size of $\mathcal{P}_{\zeta}$ and $f_{\rm{NL}}$ by many orders of magnitude. Significant changes to $\mathcal{P}_{\zeta}$ and $f_{\rm{NL}}$ from the modulus dynamics are expected in any scenario where the time-dependence of the decay rate leads to an inflaton decay that proceeds slowly or in multiple stages. 

Although our primary objective was to study the characteristics of the density perturbations generated from modulated reheating, it would be interesting to follow the fate of the modulus field after reheating and its consequences to cosmology. For example, the energy density of the modulus may increase to a non-negligible size after reheating, affecting the dynamics of the Universe. If the modulus starts oscillating around its potential minimum, the resulting isocurvature perturbations may provide important constraints when building models. We further remark that modulus particles may be produced as the inflaton decays, through the coupling term between the modulus, the inflaton, and its decay products\cite{kobayashi12}.   Moreover, this coupling term may contribute to the effective potential for the modulus, especially as the inflaton decay proceeds.  It would be interesting to study such effects in specific modulated reheating scenarios.  We have shown that these analyses must also include a dynamical modulus because the amplitude of the power spectrum and the non-Gaussianity of the density perturbations can be significantly modified by the evolution of the modulus field. It is essential to include these modifications in future investigations of modulated reheating, especially now that we have very tight constraints on the statistics of the primordial curvature perturbations.

\subsection*{Acknowledgment}
We would like to thank Jonathan Braden and Masahide Yamaguchi for valuable discussions. TK also thanks Fuminobu Takahashi for helpful conversations.  
This research was supported in part by Perimeter Institute for Theoretical Physics. Research at Perimeter Institute is supported by the Government of Canada through Industry Canada and by the Province of Ontario through the Ministry of Research and Innovation.  AE also acknowledges support from the Canadian Institute for Advanced Research Global Scholar Academy.

\bibliographystyle{JHEP} 
\bibliography{ref}
\end{document}